\documentclass[%
 aip,
 jmp,%
 amsmath,amssymb,
 reprint,%
]{revtex4-1}

\usepackage{graphicx}% Include figure files
\usepackage{dcolumn}% Align table columns on decimal point
\usepackage{bm}% bold math
\usepackage{multirow}
\usepackage{booktabs}
\usepackage{wasysym}
\usepackage{txfonts}

\newcolumntype{K}[1]{>{\centering\arraybackslash}p{#1}}

\usepackage{color}
\newcommand*{\mathcolor}{}
\def\mathcolor#1#{\mathcoloraux{#1}}
\newcommand*{\mathcoloraux}[3]{%
  \protect\leavevmode
  \begingroup
    \color#1{#2}#3%
  \endgroup
}

\usepackage{array}
\usepackage[
top    = 1.8cm,
bottom = 1.8cm,
left   = 2cm,
right  = 2cm]{geometry}

\begin{document}

%\preprint{AIP/123-QED}

\title{Fluids with competing interactions: II. Validating a free energy model for equilibrium cluster size}

\author{Jonathan A. Bollinger}
\affiliation{McKetta Department of Chemical Engineering, University of Texas at Austin, Austin, Texas 78712, USA}

\author{Thomas M. Truskett}
\email{truskett@che.utexas.edu}
\affiliation{McKetta Department of Chemical Engineering, University of Texas at Austin, Austin, Texas 78712, USA}

\date{\today}

\begin{abstract}
Using computer simulations, we validate a simple free energy model that can be analytically solved to predict the equilibrium size of self-limiting clusters of particles in the fluid state governed by a combination of short-range attractive and long-range repulsive pair potentials. The model is a semi-empirical adaptation and extension of the canonical free energy-based result due to Groenewold and Kegel [\emph{J. Phys. Chem. B}, 105 (2001)], where we use new computer simulation data to systematically improve the cluster-size scalings with respect to the strengths of the competing interactions driving aggregation. We find that one can adapt a classical nucleation like theory for small energetically-frustrated aggregates provided one appropriately accounts for a size-dependent, microscopic energy penalty of interface formation, which requires new scaling arguments. This framework is verified in part by considering the extensive scaling of intracluster bonding, where we uncover a superlinear scaling regime distinct from (and located between) the known regimes for small and large aggregates. We validate our model based on comparisons against approximately 100 different simulated systems comprising compact spherical aggregates with characteristic (terminal) sizes between six and sixty monomers, which correspond to wide ranges in experimentally-controllable parameters.
\end{abstract}

%Interestingly, intercluster effects need not be considered to obtain correct predictions even for rather non-dilute conditions.

\pacs{Valid PACS appear here}% PACS, the Physics and Astronomy
%Use showkeys class option if keyword
\maketitle

% &
% &&&
% &&&&&
% ===============================================================================================================
\section{Introduction}
% ===============================================================================================================
% &&&&&
% &&&
% &

%
%%
%%%
Over the past century, colloidal aggregation has been observed and described in a wide range of contexts via progressively more powerful experimental techniques, phenomenological frameworks, and quantitative models~\cite{Smoluchowski1916,DerjaguinLandau1941,VerweyOverbeek1948,Lin1989,Anderson2002,Israelachvili2011}. Spanning processes from droplet nucleation and growth, gel and glass formation, various self-assembly processes, etc., an overarching goal has been to use statistical mechanical or molecular thermodynamic approaches adopted from atomic systems and, if necessary, empirical rules to relate the \emph{strength} and \emph{lengthscale} of particle interactions to resulting equilibrium (or non-equilibrium) structures and the thermodynamics (and kinetics) of their formation. These types of relations, especially when based on physically-intuitive thermodynamic arguments, are not only of fundamental importance, but also highlight pathways for engineering materials at the nano- to microscopic level.

In this article, we focus on fluids where interactions between primary particles (monomers) are characterized by attractions acting at small lengthscales close to contact that \emph{compete} with repulsions acting at larger lengthscales. This class of interactions can drive the reversible formation of \emph{equilibrium cluster phases} composed of self-terminating aggregates (droplets) of monomers. Such cluster phases have been the focus of much recent work, ranging from theoretical and computational studies of idealized colloidal or nanoparticle suspensions~\cite{GroenewoldKegel2001,Sciortino2004,ArcherWilding2007,ToledanoSciortino2009,JiangWu2009,Bomont2012,Godfrin2013,GodfrinWagnerLiu2014,ManiBolhuis2014,Sweatman2014,JadrichBollinger2015,NguyenGlotzer2015,ZhuangCharbonneau2016} to experimental demonstrations for both archetypal colloidal particles~\cite{Campbell2005,Klix2010,Zhang2012,XiaGlotzer2012} and heterogeneous monomers with anisotropic interactions like proteins~\cite{Yethiraj2003,Stradner2004,PorcarLiu2010,LiuBaglioni2011,Johnston2012,ParkGlotzer2012,Yearley2014,Godfrin2016}. Despite the range of materials and lengthscales, the broad underlying formulation principles appear universal: induce (or allow) depletion (dispersion) attractions between monomers to drive aggregation while simultaneously controlling electrostatic repulsions between the ionic double-layers of monomers such that they collectively build up to attenuate growth.

While this basic paradigm of frustrating interactions is well-accepted, it is not yet established how to best describe observed cluster phases in terms of their thermodynamics and phenomenology. For example, is it possible to develop a simple and physically-motivated free-energy model which can generate accurate predictions of characteristic terminal cluster size \(N^{*}\) based on experimentally-tunable variables governing monomer interactions? We address this question here by directly comparing free energy-based predictions of such a phenomenological approach against computer simulations for one of the most approachable and idealized cluster-forming models: the short-range attractive, long-range repulsive (SALR) pair potential~\cite{Sciortino2004}. Once the behavior for this simple system that coarse-grains over many microscopic details of the short-range interactions, electrostatic double-layers, solvent, etc. is better understood, the goal is then to expand the framework to include more complex free energy contributions relevant for specific realizable colloidal suspensions.

First, we first review the canonical \emph{a priori} free-energy treatment for clustering colloidal suspensions due to Groenewold and Kegel~\cite{GroenewoldKegel2001,Groenewold2004,Zhang2012}, where we compare its predictions for cluster size \(N^{*}\) against a computational survey of phases comprising compact spherical aggregates. We take great care to clarify how this elegant and frequently-cited model adapts the classical nucleation theory of non-terminating droplets (or crystals)~\cite{Debenedetti1996,Auer2000,Sear2007} for the SALR systems of interest by treating the latter as purely-attractive reference fluids superimposed with perturbative effects due to charges on the monomers and in the suspending solvent. However, while frequently cited (and adapted for related systems, e.g., driven colloids~\cite{Mani2015}), this model has not been systematically scrutinized against a large ``test set'' of cluster phases generated by gradually varying relevant independent variables, e.g., monomer surface charge \(Z\). By conducting tests that align with the phenomenological assumptions underlying the model (e.g., apolar solvents, low cluster density), we readily find that the analytical predictive formula derived from their model exhibits a spurious scaling for the range of stable cluster sizes observable in systems governed by SALR pair potentials.

With this knowledge in-hand, we describe and validate an alternative free energy model that \emph{quantitatively} predicts the characteristic cluster size \(N^{*}\) for approximately 100 different simulated SALR systems, which comprise compact spherical aggregates in the size range \(6 \leq N^{*} \leq 60\) for wide ranges in monomer packing fraction \(\phi\), attraction strength \(\beta\varepsilon\), monomer surface \(Z\), and solvent screening length \(\kappa^{-1}/d\) (notably, even finite values outside the apolar limit). In essence, we find that a framework built on classical nucleation theory can indeed describe the thermodynamics of frustrated, finite-sized clusters provided one introduces a \emph{size-dependent} enthalpic penalty of interface formation that accounts for the missing coordination bonds of ``surface'' particles in clusters. In justifying this approach, we also examine how the number of intracluster short-range bonds scales with size; interestingly, we find a \emph{superlinear} crossover at our cluster sizes that bridges the previously-established scaling regimes for very small sizes~\cite{Arkus2009,Meng2010} (e.g., \(N^{*} \leq 9\)) and larger, bulk-like droplets. Surprisingly, we also demonstrate that intercluster effects need not be considered to obtain correct predictions even for rather non-dilute conditions.

% &
% &&&
% &&&&&
% ===============================================================================================================
\section{Methods}
% ===============================================================================================================
% &&&&&
% &&&
% &

\subsection{Model interactions}

To systematically test the performance of free energy models for predicting equilibrium cluster formation, it is invaluable to be able to (1) rapidly generate aggregate configurations that can be analyzed in depth and (2) unambiguously identify relevant free energy contributions. Thus, we consider one of the simplest and most frequently-used models known to form self-limiting aggregates: the short-range attractive (SA), long-range repulsive (LR) pair potential~\cite{Sciortino2004}. The combined SALR potential can be expressed

\begin{equation}~\label{eqn:uSALR}
\beta u^{\text{SALR}}_{i,j}(x_{i,j}) = \beta u^{\text{SA}}_{i,j}(x_{i,j}) + \beta u^{\text{LR}}_{i,j}(x_{i,j})
\end{equation}

\noindent where \(\beta = (k_{\text{B}}T)^{-1}\) (\(k_{\text{B}}\) is Boltzmann's constant and \(T\) is temperature); \(x=r/d\) is the non-dimensionalized interparticle separation; \(d\) is the characteristic particle diameter. We include the subscripts \(i\) and \(j\) to account for multiple particle types because we follow previous protocols~\cite{JadrichBollinger2015,JadrichSM2015} and examine size-polydisperse three-component mixtures that approximate colloids with 10\% size polydispersity. (This favors the formation of amorphous fluid clusters over crystalline dynamically-arrested clusters.~\cite{JadrichBollinger2015}) In this context, \(x_{i,j}\equiv x-(1/2)(i+j)(\Delta_{d}/d)\), where \(i \textnormal{ (or }j)=-1,0,1\) corresponds to small, medium, and large particles, respectively, and \(\Delta_{d}/d\) is a perturbation to particle diameter. Specifically, we study mixtures comprised of 20\% small, 60\% medium (characteristic size \(d\)), and 20\% large particles with \(\Delta_{d}=0.158d\).

The short-range attractions are expressed via a generalized (100-50) Lennard-Jones model

\begin{equation}~\label{eqn:uSA}
\beta u^{\text{SA}}_{i,j}(x_{i,j}) = 4[\beta\varepsilon+(1-2\delta_{i,j})\beta\Delta_{\varepsilon}] (x_{i,j}^{-100}-x_{i,j}^{-50})
\end{equation}

\noindent where \(\beta\varepsilon\) is the reference monomer-monomer attraction strength and \(\Delta_{\varepsilon}=0.25k_{\text{B}}T\) is an energetic perturbation to promote mixing of the polydisperse particles. Given its simplicity, the contribution of Eqn.~\ref{eqn:uSA} (similar to the contact attractions in the free energy model of Groenewold and Kegel~\cite{GroenewoldKegel2001}) does not specify the microscopic or chemical details; i.e., whether the attractions arise from depletion or other short-range interactions. Generally, the range of the attraction well is approximately \(0.10d\).

Long-range repulsions are calculated on the basis of the repulsive portion of the DLVO potential~\cite{DerjaguinLandau1941,VerweyOverbeek1948}, which approximately captures the interactions of electrostatic double-layers formed around each monomer due to (homogeneously distributed) surface charge \(Z\). This is expressed~\cite{Israelachvili2011}

\begin{equation}~\label{eqn:uLR}
\beta u^{\text{LR}}_{i,j}(x_{i,j}) = \beta A_{\text{MAX}} \dfrac{\exp{\{-(x_{i,j} - 1)/(\kappa^{-1}/d)\}}}{x_{i,j}}
\end{equation}

\noindent with

\begin{equation}~\label{eqn:maxrep}
\beta A_{\text{MAX}} = \dfrac{Z^{2}(\lambda_{\text{B}}/d)}{[1+0.5/(\kappa^{-1}/d)]^{2}} %\exp{\{-(   \}}
\end{equation}

\noindent where \(\beta A_{\text{MAX}}\) is the maximum electrostatic barrier between particles at contact, \(\kappa^{-1}/d\) is the Debye-H\"{u}ckel screening length, \(Z\) is the total surface charge per monomer, and \(\lambda_{\text{B}}/d\) is the Bjerrum length of the solvent. Crucially, this formulation neglects any long-range multi-body interactions~\cite{PandavJPCB2015,PandavLang2015}, and any charge renormalization due to counterion condensation~\cite{Manning1979,Alexander1984,Ramanathan1988,Gillespie2014} or close monomer association~\cite{ParkGlotzer2012,NguyenGlotzer2015}. As our goal here is to test how even the simplest clustering systems might be described from a free energy perspective, we reserve incorporation of these phenomena for future studies.

In using this model, we set the average monomer packing fraction \(\phi = (\pi/6)\rho d^{3}\) (where \(\rho d^{3}\) is number density), charge \(Z\), and screening length \(\kappa^{-1}/d\), and then \emph{independently} tune the attraction strength \(\beta\varepsilon\)  to drive aggregation as if varying the amount of non-interacting depletant. In terms of experimental control one can exert over repulsive contributions, this picture is somewhat idealized: to wit, tunable repulsion-controlling parameters are more realistically (though still ignoring some possible interdependence) charge \(Z\), solvent relative permittivity \(\epsilon_{\text{R}}\), and solvent ionic strength \(I\). This is because, even approximately, the screening length \(\kappa^{-1}/d = \sqrt{\epsilon_{0}\epsilon_{\text{R}}k_{\text{B}}T/(2d^{2}N_{\text{A}}e^{2}I)}\) and \(\lambda_{\text{B}}/d = e^{2}/(4d\pi\epsilon_{0}\epsilon_{\text{R}}k_{\text{B}}T)\), where \(\epsilon_{0}\) is the vacuum permittivity, \(N_{\text{A}}\) is Avogadro's number, and \(e\) is the elementary charge. For simplicity, however, we universally \emph{fix} the relative Bjerrum length \(\lambda_{\text{B}}/d\), which means electrostatic effects are set via combinations of \(Z\) and \(\kappa^{-1}/d\). 
% *** NEW (added)
With this experimental picture in mind, we also note that the repulsive strength in Eqn.~\ref{eqn:maxrep} can equivalently be written \(\beta A_{\text{MAX}} = \pi d \epsilon_{0}\epsilon_{\text{R}} \Psi_{0}^{2}/(k_{\text{B}}T)\), where \(\Psi_{0}\) is the surface potential on the monomer (often assumed to approximately equal the \(\zeta\)-potential measured via electrophoresis).

\begin{figure}
\begin{center}
  \includegraphics[resolution=300,trim={0 0 0 0},clip]{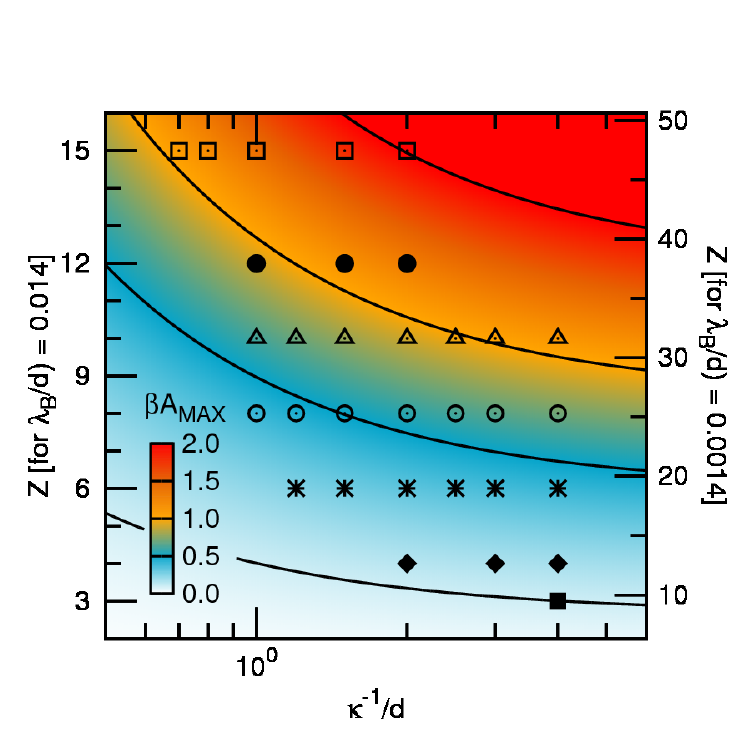}
  \caption{Maximum repulsion strength \(\beta A_{\text{MAX}} = Z^{2}(\lambda_{\text{B}}/d)/[1.0+0.5/(\kappa^{-1}/d)]^2\) plotted as a function of surface charge \(Z\) and screening length \(\kappa^{-1}/d\), where the left and right \(y\)-axes show \(Z\)-values referenced against two different reference Bjerrum lengths \(\lambda_{\text{B}}/d\). The two reference Bjerrum lengths are \(\lambda_{\text{B}}/d = 0.014\), which corresponds in real units to, e.g., \(d=50\) nm monomers in a solvent with dielectric constant \(\epsilon_{\text{R}}=80\) (equivalently, \(d=100\) nm and \(\epsilon_{\text{R}}=40\), or \(d=200\) nm and \(\epsilon_{\text{R}}=20\)); and \(\lambda_{\text{B}}/d = 0.0014\), which corresponds to \(d=500\) nm and \(\epsilon_{\text{R}}=80\) (equivalently, \(d=1\) \(\mu\)m and \(\epsilon_{\text{R}}=40\), or \(d=2\) \(\mu\)m and \(\epsilon_{\text{R}}=20\)). Symbols mark \(Z\)-(\(\kappa^{-1}/d\)) combinations tested via simulations, where Table I lists the specific combinations tested at each packing fraction \(\phi\). Throughout the manuscript, simulations are referenced by the \(Z\)-values on the left \(y\)-axis, i.e., \(Z = 3\), 4, 6, 8, 10, 12, and 15. Contours mark \(\beta A_{\text{MAX}} = 0.10\), 0.50, 1.0, and 2.0 from bottom to top.}
  \label{sch:Figure-inftyscaling}
\end{center}
\end{figure}

As illustrated in Fig. 1, we conduct a wide survey of \(Z\)-\(\kappa^{-1}/d\) combinations designed to span the the weakest repulsions that produce self-limiting aggregates (i.e., near the boundary of macrophase separation) to repulsions with strengths up to \(A_{\text{MAX}} \approx 2.0k_{\text{B}}T\). Here, note that to examine this range of repulsion strengths referenced against any plausible relative Bjerrum length \(\lambda_{\text{B}}/d\) (e.g., \(\lambda_{\text{B}}/d = 0.014\), corresponding to \(d=50\) nm monomers suspended in room temperature water with \(\lambda_{\text{B}}=0.7\) nm), one must consider monomers with \emph{very low} effective charge density. Throughout the publication, we reference \(Z\)-values based on the choice of \(\lambda_{\text{B}}/d = 0.014\), though choosing a different reference \(\lambda_{\text{B}}/d\) simply renormalizes the range of \(Z\) under consideration, with an example of this rescaling given in Fig. 1. All of the parameter combinations (\(\phi\), \(Z\), \(\kappa^{-1}/d\)) we examine are listed in Table I by their respective critical attraction strengths (discussed below). Finally, note that throughout the remainder of the publication, we notate \(\beta u^{\text{SALR}}_{i,j}(x_{i,j})\) as \(\beta u(r)\) for aesthetic simplicity unless otherwise indicated.

%%%%%%%%%%%%%%%
%%%%%%%%%%%%%%%
%%%%%%%%%%%%%%%
\subsection{Molecular dynamics simulations}~\label{lbl:sims}

We generate configurations of cluster phases via three-dimensional MD simulations of the ternary SALR mixtures described above, where we generate trajectories using LAMMPS~\cite{Plimpton1995}. We perform simulations in the NVT ensemble with periodic boundary conditions using an integration time-step of \(dt=0.001\sqrt{d^2m/(k_{\text{B}}T)}\) (taking the mass \(m=1\)) and fix temperature via a Nos\'{e}-Hoover thermostat with time-constant \(\tau=2000dt\). As outlined in Table I, we consider many combinations of charge \(Z\) and screening length \(\kappa^{-1}/d\) at four different packing fractions: \(\phi = 0.015\), 0.030, 0.060, and 0.120 (where we simulate \(N_{\text{box}} = 1920\), 2960, 6800, and 6800 particles, respectively). 
% *** NEW (added)
Beginning with randomized initial configurations, we equilibrate systems at \(\phi = 0.015\), 0.030, 0.060, and 0.120 for \(3 \text{x} 10^{7}\), \(1 \text{x} 10^{7}\), \(3 \text{x} 10^{6}\), and \(2 \text{x} 10^{6}\) steps, respectively, and confirm that they are equilibrated on the basis of energy convergence and visualization, where the latter shows that the systems are ergodic (aggregates undergo frequent intra- and intercluster rearrangements and exchanges). 
We cut-off the pair potential for a given \(Z\) and \(\kappa^{-1}/d\) such that the interaction strength at distance \(x^{\text{c}}_{i,j}\) (note explicit use of the mixture notation) is \(\beta u_{i,j}(x^{\text{c}}_{i,j}) \leq 2\text{e}^{-3}\) and the force is simultaneously \(-\text{d}[\beta u_{i,j}(x^{\text{c}}_{i,j})]/\text{d}x_{i,j} \leq 1\text{e}^{-3}\).

\begin{table}[]
\centering
\caption{Critical attraction strengths \(\beta\varepsilon^{*}\) determined from MD simulations at various \(\phi\) as a function of surface charge \(Z\) and screening length \(\kappa^{-1}/d\). Conditions with listed  \(\beta\varepsilon^{*}\) values are those used for our analysis and discussion. Symbols below the \(Z\) values correspond to those used in Figs. 2-7 (symbols are kept constant for various \(\kappa^{-1}/d\)). Note that maximum repulsion strengths \(\beta A_{\text{MAX}}\) (see Eqn.~\ref{eqn:maxrep}) are calculated based on a reference relative Bjerrum length of \(\lambda_{\text{B}}/d = 0.014\).}
\label{tbl:pf0015}
\begin{tabular}{|K{0.4cm} K{0.70cm} K{0.7cm} K{0.7cm} K{0.7cm} K{0.7cm} K{0.7cm} K{0.7cm} K{0.7cm}|}
\multicolumn{9}{l}{} \\ \hline%\hline
 \multicolumn{2}{|c|}{\multirow{3}{*}{\(\kappa^{-1}/d\)}}  & \multicolumn{7}{c|}{Z}        \\  
 & \multicolumn{1}{l|}{ } & 3 & 4 & 6 & 8 & 10 & 12 & 15 \\ %\cline{1-9}  
 & \multicolumn{1}{l|}{ } & \({\blacksquare}\) & \({\Diamondblack}\) & \({\divideontimes}\) & \({\Circle}\) & \({\vartriangle}\) & \({\CIRCLE}\) & \({\square}\) \\ \cline{1-9} 
\multicolumn{1}{|c}{\multirow{9}{*}{\rotatebox[origin=c]{90}{\(\phi=0.015\) \(\mathcolor[rgb]{0.024,0.643,0.792}{\blacksquare}\)}}} & \multicolumn{1}{l|}{0.7} & -  &  - &  - & -  & -   & -   &  6.55  \\
\multicolumn{1}{|l}{}                   & \multicolumn{1}{l|}{0.8}  & - & - & - & - & - & - &  6.80  \\
\multicolumn{1}{|l}{}                   & \multicolumn{1}{l|}{1.0} & -  & -  & -  &  5.55  &  6.00  &  6.40  &  7.10  \\
\multicolumn{1}{|l}{}                   & \multicolumn{1}{l|}{1.2} &  - &  - &  - &  5.65  &  6.10  &  -  &  -  \\
\multicolumn{1}{|l}{}                   & \multicolumn{1}{l|}{1.5} &  - & -  &  5.35  &  5.80  &  6.30  &  6.80  &  -  \\
\multicolumn{1}{|l}{}                   & \multicolumn{1}{l|}{2.0} &  - &  5.05  &  5.50  &  5.95  &  6.45  &  7.00  &  7.90  \\ 
\multicolumn{1}{|l}{}                   & \multicolumn{1}{l|}{2.5} & -  &  - &  5.55  &  6.00  &  6.60  &  -  &   - \\ 
\multicolumn{1}{|l}{}                   & \multicolumn{1}{l|}{3.0} & -  &  5.10  &  5.55  &  6.05  &  6.60  &  -  &  -  \\ 
\multicolumn{1}{|l}{}                   & \multicolumn{1}{l|}{4.0} &  4.95  &  5.10  &  5.60  &  6.10  &  6.65  & -   &  -  \\ \hline
\multicolumn{9}{l}{} \\ \hline %\hline
\multicolumn{1}{|c}{\multirow{9}{*}{\rotatebox[origin=c]{90}{\(\phi=0.030\) \(\mathcolor[rgb]{1.000,0.647,0.000}{\blacksquare}\)}}} & \multicolumn{1}{l|}{0.7} & -  &  - &  - & -  & -   & -   &  6.30  \\
\multicolumn{1}{|l}{}                   & \multicolumn{1}{l|}{0.8}  & - & - & - & - & - & - &  -  \\
\multicolumn{1}{|l}{}                   & \multicolumn{1}{l|}{1.0} & -  & -  & -  &  5.30  &  5.70  &  6.15  &  6.75  \\
\multicolumn{1}{|l}{}                   & \multicolumn{1}{l|}{1.2} &  - &  - &  - &  5.45  &  5.80  &  -  &  -  \\
\multicolumn{1}{|l}{}                   & \multicolumn{1}{l|}{1.5} &  - & -  &  5.15  &  5.55  &  5.95  &  6.45  &  -  \\
\multicolumn{1}{|l}{}                   & \multicolumn{1}{l|}{2.0} &  - &  4.80  &  5.20  &  5.65  &  6.10  &  6.55  &  7.25  \\ 
\multicolumn{1}{|l}{}                   & \multicolumn{1}{l|}{2.5} & -  &  - &  5.20  &  5.70  &  6.20  &  -  &   - \\ 
\multicolumn{1}{|l}{}                   & \multicolumn{1}{l|}{3.0} & -  &  4.90  &  5.25  &  5.70  &  6.20  &  -  &  -  \\ 
\multicolumn{1}{|l}{}                   & \multicolumn{1}{l|}{4.0} &  4.70  &  4.90  &  5.30  &  5.70  &  6.20  & -   &  -  \\ \hline
\multicolumn{9}{l}{} \\ \hline%\hline
\multicolumn{1}{|c}{\multirow{9}{*}{\rotatebox[origin=c]{90}{\(\phi=0.060\) \(\mathcolor[rgb]{0.901,0.380,0.000}{\blacksquare}\)}}} & \multicolumn{1}{l|}{0.7} & -  &  - &  - & -  & -   & -   &  6.00  \\
\multicolumn{1}{|l}{}                   & \multicolumn{1}{l|}{0.8}  & - & - & - & - & - & - &  -  \\
\multicolumn{1}{|l}{}                   & \multicolumn{1}{l|}{1.0} & -  & -  & -  &  5.00  &  5.40  &  5.65  &  6.25  \\
\multicolumn{1}{|l}{}                   & \multicolumn{1}{l|}{1.2} &  - &  - &  4.75 &  5.10  &  5.45  &  -  &  -  \\
\multicolumn{1}{|l}{}                   & \multicolumn{1}{l|}{1.5} &  - & -  &  4.80  &  5.15  &  5.50  &  5.80  &  -  \\
\multicolumn{1}{|l}{}                   & \multicolumn{1}{l|}{2.0} &  - &  4.55  &  4.85  &  5.20  &  5.55  &  5.80  &  6.40  \\ 
\multicolumn{1}{|l}{}                   & \multicolumn{1}{l|}{2.5} & -  &  - &  4.90  &  5.20  &  5.60  &  -  &   - \\ 
\multicolumn{1}{|l}{}                   & \multicolumn{1}{l|}{3.0} & -  &  4.60  &  4.85  &  -  &  5.60  &  -  &  -  \\ 
\multicolumn{1}{|l}{}                   & \multicolumn{1}{l|}{4.0} &  4.40  &  4.60  &  4.85  &  5.20  &  5.60  & -   &  -  \\ \hline
\multicolumn{9}{l}{} \\ \hline%\hline
\multicolumn{1}{|c}{\multirow{9}{*}{\rotatebox[origin=c]{90}{\(\phi=0.120\) \(\mathcolor[rgb]{1.000,0.000,0.000}{\blacksquare}\)}}} & \multicolumn{1}{l|}{0.7} & -  &  - &  - & -  & -   & -   &  5.20  \\
\multicolumn{1}{|l}{}                   & \multicolumn{1}{l|}{0.8}  & - & - & - & - & - & - &  5.20  \\
\multicolumn{1}{|l}{}                   & \multicolumn{1}{l|}{1.0} & -  & -  & -  &  -  &  -  &  4.95  &  5.20  \\
\multicolumn{1}{|l}{}                   & \multicolumn{1}{l|}{1.2} &  - &  - &  - &  -  & -  &  -  &  -  \\
\multicolumn{1}{|l}{}                   & \multicolumn{1}{l|}{1.5} &  - & -  &  -  &  -  &  4.75  &  4.95  &  5.20  \\
\multicolumn{1}{|l}{}                   & \multicolumn{1}{l|}{2.0} &  - &  -  &  -  &  4.60  &  4.75  &  4.95  &  -  \\ 
\multicolumn{1}{|l}{}                   & \multicolumn{1}{l|}{2.5} & -  &  - &  -  &  4.60  &  -  &  -  &   - \\ 
\multicolumn{1}{|l}{}                   & \multicolumn{1}{l|}{3.0} & -  &  -  &  -  &  -  &  -  &  -  &  -  \\ 
\multicolumn{1}{|l}{}                   & \multicolumn{1}{l|}{4.0} &  -  & -  &  -  &  -  &  -  & -   &  -  \\ \hline
\end{tabular}
\end{table}

To make our analysis tractable, we focus on states corresponding with the \emph{onset of clustering} (i.e., at the cluster transition locus), where the phases are composed of fluid aggregates with characteristic size \(N^{*}\), but the systems have not yet begun to form percolated phases or become dynamically arrested. To characterize the size of equilibrium aggregates, we calculate cluster-size distributions (CSDs), which quantify the probability \(p(N)\) of observing clusters comprising \(N\) particles. Here, we follow the established convention~\cite{Sciortino2004,GodfrinWagnerLiu2014,ManiBolhuis2014,JadrichBollinger2015} of considering two monomers part of the same cluster if they are directly bonded to one another (i.e., within the range of the attractive well) or each directly bonded to a shared neighbor (i.e., are connected via some percolating pathway).

In turn, to locate the cluster transition locus, we make sweeps in attraction strength \(\beta\varepsilon\) (at increments of \(\Delta\varepsilon=0.05k_{\text{B}}T\)) and identify states at the onset of clustering based on the following criteria: (1) the \(p(N)\) distribution exhibits a visibly-apparent local maximum (mode) at some \(1 < N^{*} \ll N_{\text{box}}\), where the corresponding local minimum between \(N=1\) and \(N^{*}\) is notated as \(N_{\text{min}}\); and (2) that 80\% of the particles in the system participate in aggregates of size \(N \geq N_{\text{min}}\), i.e., \(0.80 = \sum_{n=N_{\text{min}}}^{N_{\text{box}}} p(N)\) where \(p(N)\) is appropriately normalized. Taken together, these conditions correspond to the emergence of meaningful bimodality (coexistence) in \(p(N)\) between \(N = 1\) and the cluster mode \(N^{*}\). In this way, we obtain the characteristic cluster size \(N^{*}\) associated with a particular combination of \(\phi\), \(Z\), and \(\kappa^{-1}/d\) and the corresponding \emph{critical} attraction strength \(\beta\varepsilon^{*}\). All of the parameter combinations we consider in our analysis are listed by their respective \(\beta\varepsilon^{*}\) values in Table I.

% &
% &&&
% &&&&&
% ===============================================================================================================
\section{Results \& Discussion}
% ===============================================================================================================
% &&&&&
% &&&
% &

\subsection{Observed cluster sizes and shapes in simulations}~\label{sect:shape}

Before discussing free energy models for characteristic cluster size \(N^{*}\), we begin by briefly describing the cluster morphologies under examination: for the approximately 100 different combinations of packing fraction \(\phi\), surface charge \(Z\), and screening length \(\kappa^{-1}/d\) that we consider (listed in Table I), we observe phases at the corresponding critical attraction strengths \(\beta\varepsilon^{*}\) that comprise compact spherical clusters with characteristic sizes in the range \(6 \leq N^{*} \leq 60\), as plotted in Fig. 2. In terms of cluster shape, we find that by measuring the radius of gyration \(R_{\text{G}}/d\) and plotting it versus cluster size \(N^{*}\), our results obey the relation

\begin{equation}~\label{eqn:Rgscaling}
R_{\text{G}}/d = \alpha(\phi) N^{*(1/d_{\text{f}})} \text{  with  } d_{\text{f}} = 3
\end{equation}

\noindent where \(\alpha(\phi)\) is a \(\phi\)-dependent prefactor of magnitude approximately \(1/2\) (hereafter notated \(\alpha\)). Together with the fractal dimension \(d_{\text{f}} = 3\), this signifies that the aggregates are compact objects, and visual inspection of the MD trajectories confirms the clusters are indeed highly-packed amorphous droplets that are spherical on average and undergo frequent intracluster rearrangement and intercluster exchange (seen previously~\cite{JadrichBollinger2015,JadrichSM2015}). As shown in the inset of Fig. 2, the clusters do become slightly less packed with increasing \(\phi\), which is attributable to an increasing frequency of intercluster exchange. (These transfer events tend to instantaneously but, on average, isotropically distort the clusters, effectively expanding them.) We discuss trends in cluster size and shape from a different perspective (and in more detail) in the accompanying publication.

% !!!
\begin{figure}
\begin{center}
  \includegraphics[trim={0 0 0 0},clip]{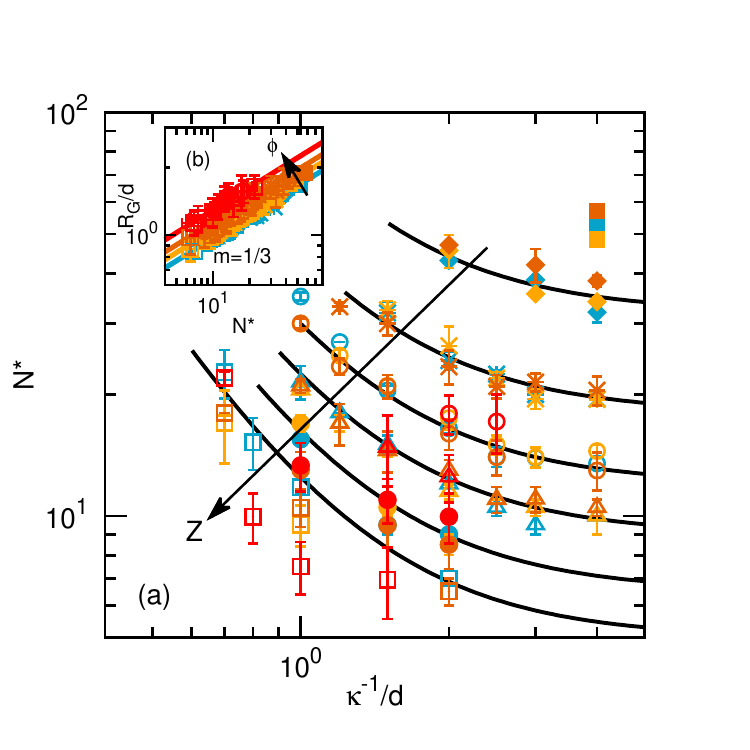}
  \caption{(a) Measured cluster size \(N^{*}\) versus screening length \(\kappa^{-1}/d\) for all \(\phi\), \(Z\), and \(\kappa^{-1}/d\) combinations tested. Blue, yellow, orange, and red symbols correspond to measurements from simulations at \(\phi = 0.015\), 0.030, 0.060, and 0.120, respectively. Contours are guides to the eye for constant \(Z\): from top to bottom, \(Z=3.0\) (no line), 4.0, 6.0, 8.0, 10.0, 12.0, and 15.0. These contours are plotted according to the formula \(N^{*}/N^{\text{est}}_{\infty} = 1.0 + 1.5/(\kappa^{-1}/d)^2\), where \(N^{\text{est}}_{\infty}\) is the estimated cluster size in the Coulombic limit (i.e., \(\kappa^{-1}/d \rightarrow \infty\)). (b) Cluster radius of gyration \(R_{\text{G}}/d\) versus characteristic cluster size \(N^{*}\), both measured from MD simulations. Lines are empirical fits of the form \(R_{\text{G}}/d = \alpha N^{*1/3}\), where \(\alpha\) is a dimensionless prefactor corresponding to \(\alpha = 0.45\), 0.49, 0.53, and 0.60 for \(\phi= 0.015\), 0.030, 0.060, and 0.120, respectively. Symbol types in (a) and (b) correspond to constant charge \(Z\) as listed in Table I (note that we test various screening lengths \(\kappa^{-1}/d\) at each \(Z\)).}
  \label{sch:Figure-inftyscaling}
\end{center}
\end{figure}

In terms of cluster number size \(N^{*}\), there are two important observations from Fig. 2: (1) characteristic cluster size depends only weakly on packing fraction for the range of \(0.015 \leq \phi \leq 0.120\); and (2) the morphologies associated with \emph{unscreened} electrostatic repulsions (i.e., \(\kappa^{-1}/d \rightarrow \infty\)) are effectively generated when the screening length approaches \(\kappa^{-1}/d \approx 4.0\). As shown by considering Figs. 2(a) and (b) simultaneously, increasing packing fraction \(\phi\) (given fixed \(Z\) and \(\kappa^{-1}/d\)) does not systematically shift \(N^{*}\), but does slightly inflate the cluster radius \(R_{\text{G}}/d\). (We do note that the CSD peaks at \(N^{*}\) also become wider with increasing \(\phi\) due to more frequent intercluster contacts.) The second point is apparent based on Fig. 2(a), which demonstrates that for the larger screening lengths \(\kappa^{-1}/d\) tested, cluster sizes \(N^{*}\) at fixed \(\phi\) and \(Z\) have already nearly reached \emph{asymptotic} values, i.e.,

\begin{equation}
\lim_{\kappa^{-1}/d \rightarrow \infty} N^{*}_{\infty} \approx N^{*} \text{ at } \kappa^{-1}/d = 4.0
\end{equation}

\noindent This ability to access the Coulombic limit at finite \(\kappa^{-1}/d\) is important for the following sections.

%%%%%%%%%%%%%%%
%%%%%%%%%%%%%%%
%%%%%%%%%%%%%%%
\subsection{Existing free energy model for cluster size}~\label{sect:Groenewold}

We now begin our discussion of the canonical framework for cluster formation due to Groenewold and Kegel~\cite{GroenewoldKegel2001} (with subsequent follow-ups~\cite{Groenewold2004,Zhang2012}), with an emphasis on making clear important concepts and assumptions underpinning the model. The model aims to predict characteristic cluster size \(N^{*}_{\infty}\) for large and perfectly monodisperse aggregates governed by short-range attractions (SA) and long-range (LR) unscreened Coulombic interactions between monomers (the subscript alludes to the \(\kappa^{-1}/d \rightarrow \infty\) limit). This prediction necessarily begins with an expression for the extensive free energy \(\beta \Delta F\) of cluster formation as a function of \(N\) (agnostic to \(N^{*}_{\infty}\)): 

\begin{equation}~\label{eqn:start}
\beta \Delta F = \beta F_{N} - N \beta F_{1}
\end{equation}

\noindent where \(\beta F_{N}\) and \(\beta F_{1}\) are the free energies of the \(N\)-sized clusters and monomers, respectively.

The free energy change \(\beta \Delta F\) is broken into \emph{reference} and \emph{perturbative} contributions: the reference portion is taken to be the free energy of aggregate formation for a SA (i.e., purely attractive) fluid, which can be described via the classical nucleation theory (CNT) for large droplets (or crystals)~\cite{Debenedetti1996,Auer2000,Sear2007}. Meanwhile, the perturbations are any contributions to the free energy due to the electrostatic effects. This is simply expressed:

\begin{equation}~\label{eqn:start}
\beta \Delta F = \beta \Delta F^{\text{SA}} + \beta \Delta F^{\text{LR}}
\end{equation}

\noindent where we detail these (reference) attractive and (perturbative) repulsive free energy differentials in order below.

The CNT-based free energy contributions of the reference SA system comprise two terms, which capture competing effects that scale with aggregate volume and surface area, respectively. The first term accounts for the transfer of monomers from the low-density dispersed phase to the dense (bulk) fluid or crystal phase corresponding to the cluster interior. This transfer is characterized by a favorable change in chemical potential per particle with the magnitude \(\beta\Delta\mu_{0}^{\text{SA}}\). The second term is an enthalpic penalty~\cite{SearJCP1999,SearPRE1999} characterized by surface tension \(\beta\gamma^{\text{SA}}d^{2}\), which accounts for the relative number of ``missing'' intracluster coordination bonds \(z_{\text{c,m}}\) of the particles at the droplet surface relative to, e.g., the bulk-like coordination number \(z_{\text{c,0}}\) of the cluster interior~\footnote{In principle, the free-energy penalty also includes an \emph{entropic} contribution due to the increased mobility particles might have at the droplet surface compared to the droplet interior; however, this contribution is often negligible~\cite{SearPRE1999}. Groenewold and co-workers do not address this issue~\cite{GroenewoldKegel2001,Groenewold2004,Zhang2012}, but for our systems, where clusters possess fluid-like structures with frequent rearrangement between interior to exterior (nevermind frequent intercluster exchange), we also expect this entropic differential to be small.}. These contributions can be written

\begin{equation}~\label{eqn:CNTterms}
\beta \Delta F^{\text{SA}} = -N\beta\Delta\mu_{0}^{\text{SA}} + 4\pi(R_{\text{c}}/d)^{2} (\beta\gamma^{\text{SA}}d^{2})
\end{equation}

\noindent where, reflecting our observed morphologies, we incorporate the expression for cluster surface area assuming \emph{spherical} droplets with radius \(R_{\text{c}}/d\). Going forward, this radius is considered interchangeable with the radius of gyration within some \(O(1)\) prefactor, i.e., \(R_{\text{c}} \approx R_{\text{G}}\).

In turn, the perturbative electrostatic contributions are treated as arising from unscreened repulsions acting between all intracluster pairs of particles (i.e., \(N(N-1)/2 \approx N^{2}/2\) interactions), which can be written: 

\begin{equation}~\label{eqn:Kegelrepulsion}
\beta \Delta F^{\text{LR}} \approx \dfrac{\langle \beta u^{\text{LR}}\rangle N^{2}}{2} \approx \dfrac{Z^{2}(\lambda_{\text{B}}/d)N^{2}}{2(R_{\text{c}}/d)} 
\end{equation}

\noindent where \(\langle \beta u^{\text{LR}} \rangle \approx Z^{2}(\lambda_{\text{B}}/d)/(R_{\text{c}}/d)\) is the Coulombic limit (\(\kappa^{-1}/d \rightarrow \infty\)) of the DLVO-type potential of Eqns.~\ref{eqn:uLR} and ~\ref{eqn:maxrep} evaluated at \(r = R_{\text{c}}/d\), which assumes that the characteristic (average) intracluster pair distance is simply the cluster radius~\footnote{Zhang and co-workers~\cite{Zhang2012} report the wrong exponent with respect to \(N\) for this term.}. The form of Eqn.~\ref{eqn:Kegelrepulsion} implies that the repulsive free-energy contribution of each monomer in the dispersed phase is truly negligible compared to the intracluster contribution, which is consistent with the choice of Groenewold and Kegel to \emph{ignore intercluster interactions}, i.e., consider the limit of very low \(\phi\). Note that Groenewold and Kegel also originally include a term (see Eqn. 18 in Ref.~\citenum{GroenewoldKegel2001}) that roughly accounts for counterion condensation~\cite{Manning1979,Alexander1984,Ramanathan1988,Gillespie2014}, which could occur for strong bare surface charges. However, we neglect this contribution because their approximation naturally drops out of the subsequent analysis and the coarse-grained SALR potential considered here only captures a constant net-effective charge.

Given these expressions for the free energy contributions, one can proceed to the crux of the analysis: identifying the characteristic cluster size \(N^{*}_{\infty}\) at which the driving force to associate per monomer is at its largest magnitude (or energetic minimum), i.e., \(\beta\Delta f(N^{*}) \equiv \min_{N} [\beta\Delta f(N)]\) where \(\beta\Delta f(N) \equiv \beta\Delta F(N)/N\). Of course, here one requires a \(\beta\Delta f(N)\) function where the sole dependent variable is \(N\). By combining Eqns.~\ref{eqn:CNTterms} and ~\ref{eqn:Kegelrepulsion} with the known relation between cluster radius and number size \(R_{\text{G}}/d = \alpha N^{1/3}\) for compact spherical aggregates, one can readily write:

\begin{equation}~\label{eqn:Kegelintensive}
\beta \Delta f(N) = -\beta\Delta\mu_{0}^{\text{SA}} + \dfrac{4\pi\alpha^{2}(\beta\gamma^{\text{SA}}d^{2})}{N^{1/3}} + \dfrac{Z^{2}(\lambda_{\text{B}}/d)N^{2/3}}{2\alpha} 
\end{equation}

\noindent and evaluate its derivative to find the global minimum

\begin{equation}~\label{eqn:Kegelequality}
\dfrac{\text{d}(\beta \Delta f)}{\text{d}N}\bigg|_{N^{*}_{\infty}} = 0 = -\dfrac{4\pi\alpha^{2}(\beta\gamma^{\text{SA}}d^{2})}{3{N^{*}_{\infty}}^{4/3}} + \dfrac{Z^{2}(\lambda_{\text{B}}/d)}{3\alpha{N^{*}_{\infty}}^{1/3}} 
\end{equation}

\noindent which, dropping prefactors, gives the scaling relation: %provides the central result:

\begin{equation}~\label{eqn:Kegelsurf}
N^{*}_{\infty} \propto \dfrac{\beta\gamma^{\text{SA}}d^{2}}{Z^{2}(\lambda_{\text{B}}/d)}
\end{equation}

\noindent This states that cluster size is simply governed by the strength of the surface energy relative to the characteristic strength of electrostatic repulsion.

To write Eqn.~\ref{eqn:Kegelsurf} completely in terms of experimentally tunable parameters, one then approximates~\cite{SearJCP1999,SearPRE1999,Zhang2012} 
the surface tension of the SA reference fluid \(\beta\gamma^{\text{SA}}d^{2}\) as scaling like the attraction strength \(\beta\varepsilon\) multiplied by the aforementioned number of missing bonds per surface particle \(z_{\text{c,m}}\) (divided by a ``surface area'' per monomer \(A_{\text{m}}\)), i.e., 

\begin{equation}~\label{eqn:Zhangzm}
\beta \gamma^{\text{SA}} d^{2} \approx \dfrac{z_{\text{c,m}}\beta\varepsilon}{(A_{\text{m}}/d^{2})}
\end{equation}

\noindent Because \(z_{\text{c,m}}\) is considered constant with respect to \(N\) for large, low-curvature droplets, combining Eqns.~\ref{eqn:Kegelsurf} and ~\ref{eqn:Zhangzm} leads to the master \emph{a priori} scaling relation 

\begin{equation}~\label{eqn:Kegelmaster}
N^{*}_{\infty} \propto \dfrac{\beta\varepsilon}{Z^{2}(\lambda_{\text{B}}/d)}
\end{equation}

\noindent Reintroducing prefactors, Eqn.~\ref{eqn:Kegelmaster} is written \(N^{*}_{\infty} = \alpha\nu_{0}\beta\varepsilon/[Z^{2}(\lambda_{\text{B}}/d)]\), where \(\alpha\) remains from the repulsive term in Eqn.~\ref{eqn:Kegelintensive}, and \(\nu_{0}\) is a prefactor that is the product of \(z_{\text{c,m}}\) and some conversion factor to arrive at a surface energy per area.

%%%%%%%%%%%%%%%
%%%%%%%%%%%%%%%
%%%%%%%%%%%%%%%
\subsection{Observed size-scaling in the Coulombic limit}~\label{sect:rep}

Given our wide survey of compact spherical cluster morphologies, we can perform the first \emph{systematic} test of the master scaling law given by Eqn.~\ref{eqn:Kegelmaster} for SALR pair potentials by plotting measured cluster sizes \(N^{*}\) for systems with sufficiently large screening lengths \(\kappa^{-1}/d\) at various \(\phi\), \(Z\), and \(\beta\varepsilon\). Specifically, in Fig. 3, we plot \(N^{*}\) values observed at critical attraction strengths \(\beta\varepsilon^{*}\) and screening length \(\kappa^{-1}/d = 4.0\), where the latter corresponds to effectively unscreened systems (see Section~\ref{sect:shape}) as assumed in writing Eqn.~\ref{eqn:Kegelmaster}. Here, we note that we use the version of Eqn.~\ref{eqn:Kegelmaster} that incorporates prefactors \(\alpha\) and \(\nu_{0}\), which shift predicted sizes approximately in line with the measured \(N^{*}\) values (of course, including or excluding these prefactors does not affect scaling itself).

% \(\nu_{0} \approx 3.40\)
\begin{figure}
\begin{center}
  \includegraphics[trim={0 0 0 0},clip]{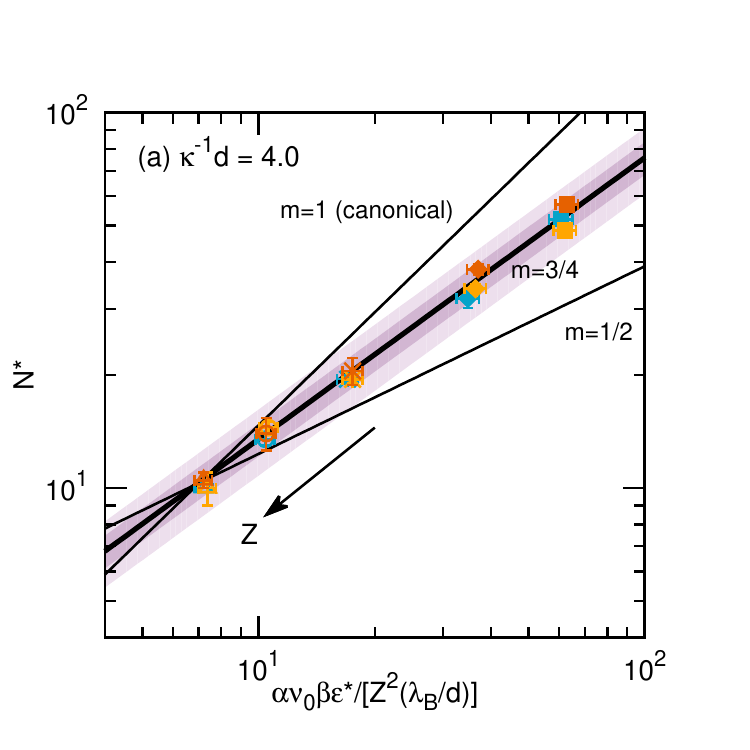}
  \caption{(a) Measured cluster size \(N^{*}_{\infty}\) in the Coulombic limit (approximated by systems with \(\kappa^{-1}/d = 4.0\)) versus the master scaling ratio of Eqn.~\ref{eqn:Kegelmaster} plotted using measured critical attraction strengths \(\beta\varepsilon^{*}\) and corresponding characteristic repulsion strengths \(Z^{2}(\lambda_{\text{B}}/d)\). Blue, yellow, and orange symbols correspond to measurements from simulations at \(\phi = 0.015\), 0.030, and 0.060, respectively, for charges \(Z=3.0\), 4.0, 6.0, 8.0, and 10.0 (top to bottom). Thick black line corresponds to the empirical scaling of Eqn.~\ref{eqn:empscaling} with exponent of \(m=3/4\) (i.e., \(N^{*}_{\infty} \propto \{\beta\varepsilon^{*}/[Z^{2}(\lambda_{\text{B}}/d)]\}^{3/4}\)) and dark (light) purple shadings correspond to 10\% (20\%) deviation from this scaling. Thin black lines show scalings for alternate exponents, where the \(m=1\) scaling (see Eqn.~\ref{eqn:Kegelmaster}) derives from the canonical free energy model of Groenewold and Kegel~\cite{GroenewoldKegel2001,Groenewold2004,Zhang2012}. Note that in this figure, we plot predicted cluster sizes (\(x\)-axis) based on including the \(\phi\)-dependent prefactor \(\alpha\) for the radius of gyration (see Fig. 2) and the (here, arbitrary) constant prefactor \(\nu_{0} \approx 3.40\) (see text). Symbol types correspond to constant charge \(Z\) as listed in Table I (note that we test various screening lengths \(\kappa^{-1}/d\) at each \(Z\)).}
  \label{sch:Figure-inftyscaling}
\end{center}
\end{figure}

In Fig. 3, we do indeed observe a master \(\phi\)-independent relation between the \(N^{*}\) values measured in simulations and the relative strength of attractions and repulsions between monomers, i.e., the ratio \(\beta\varepsilon/[Z^{2}(\lambda_{\text{B}}/d)]\); however, the observed scaling \emph{does not reflect the exponent of 1} that is expected based on the free energy model underlying Eqn.~\ref{eqn:Kegelmaster}. 
Instead, we clearly observe the empirical relation

\begin{equation}~\label{eqn:empscaling}
N^{*}_{\infty} \propto \Bigg[\dfrac{\beta\varepsilon}{Z^{2}(\lambda_{\text{B}}/d)}\Bigg]^{3/4}
\end{equation}

\noindent for various cluster sizes and packing fractions. This immediately begs the questions: what \emph{alternative} (and, ideally, comparatively simple) free energy model for SALR systems results in this softer master scaling? and furthermore, can this alternative model readily predict \(N^{*}\) for \emph{finite} screening lengths \(\kappa^{-1}/d\)?

To ascertain what new model can capture the empirically-observed scaling in Fig. 3 (and be extended for generic \(\kappa^{-1}/d\)), we first ought to identify which of the current free energy terms in Eqns.~\ref{eqn:CNTterms} and ~\ref{eqn:Kegelrepulsion} correctly (or incorrectly) describe the energetics of cluster formation in the MD simulations. Given its simplicity, the most straightforward candidate to consider is the \emph{repulsive} free energy contribution of Eqn.~\ref{eqn:Kegelrepulsion}, which we can test against MD configurations by adding up the total repulsive energies (between all intracluster pairs of monomers) of simulated clusters as a function of characteristic size \(N^{*}\).

% FIGURE 2
\begin{figure}
\begin{center}
  \includegraphics[trim={0 0 0 0},clip]{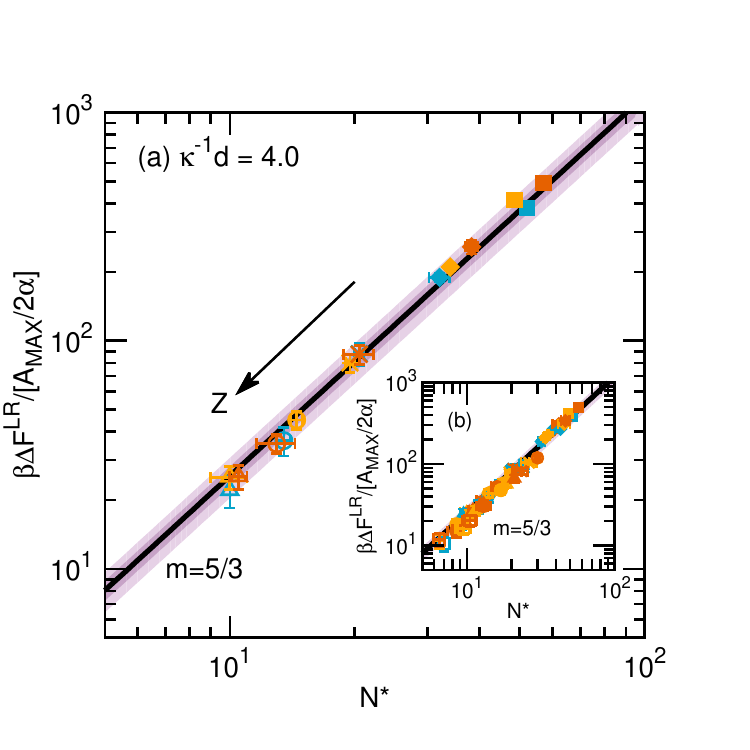}
  \caption{(a) Total intracluster repulsion energy \(\beta \Delta F^{\text{LR}}\) scaled by maximum repulsion barrier \(\beta A_{\text{MAX}} = Z^{2}(\lambda_{\text{B}}/d)/[1.0+0.5/(\kappa^{-1}/d)]^2\) and \(\phi\)-dependent prefactor \(\alpha\) for the radius of gyration (see Fig. 2), plotted versus cluster size \(N^{*}\) for \(Z=3.0\), 4.0, 6.0, 8.0, and 10.0 (top to bottom) and \(\kappa^{-1}/d = 4.0\) (effectively \(\kappa^{-1}/d \rightarrow \infty\)). Blue, yellow, and orange symbols correspond to measurements from simulations at \(\phi = 0.015\), 0.030, and 0.060, respectively. (b) Same but for \(\kappa^{-1}/d = 4.0\), 3.0, 2.0, and 1.0 at all correspondingly tested \(Z\) values (see Table I). For (a) and (b), thick black line corresponds to the expression \(\beta \Delta F^{\text{LR}}/[\beta A_{\text{MAX}}/(2\alpha)] = N^{*5/3}\) and dark (light) purple shadings correspond to 10\% (20\%) deviation from this scaling. Symbol types in (a) and (b) correspond to constant charge \(Z\) as listed in Table I (note that we test various screening lengths \(\kappa^{-1}/d\) at each \(Z\)).}
  \label{sch:Figure-inftyscaling}
\end{center}
\end{figure}

As shown in Fig. 4, we observe that the repulsive free energy contribution of Eqn.~\ref{eqn:Kegelrepulsion} quantitatively describes MD results in the unscreened limit and, with a simple extension, also works for finite screening lengths \(\kappa^{-1}/d\); in other words, the current perturbative free energy term capturing electrostatics is \emph{self-consistent} and should be retained. In Fig. 4(a), we see that \(\beta \Delta F^{\text{LR}}\) measured in simulations, when normalized by the maximum repulsion barrier \(\beta A_{\text{MAX}} = Z^{2}(\lambda_{\text{B}}/d)\) (corresponding to the \(\kappa^{-1}/d \rightarrow \infty\) limit of Eqn.~\ref{eqn:maxrep}), scales as \(N^{5/3}\). Of course, this \(N^{5/3}\) scaling is expected given \(N^{2}\) intracluster pair interactions occurring on the lengthscale of the cluster radius, which scales as \(N^{1/3}\) (see Eqn.~\ref{eqn:Kegelrepulsion}. Meanwhile, Fig. 2(b) demonstrates that the same scaling holds for finite \(\kappa^{-1}/d\) away from the Coulombic limit provided one appeals to the more generalized form of Eqn.~\ref{eqn:maxrep} for the maximum repulsive barrier energy, i.e., \(\beta A_{\text{MAX}} = Z^{2}(\lambda_{\text{B}}/d)/[1.0+0.5/(\kappa^{-1}/d)]^2\).

%%%%%%%%%%%%%%%
%%%%%%%%%%%%%%%
%%%%%%%%%%%%%%%
%\subsection{Surface effects at ``intermediate'' cluster size}~\label{sect:surf}
\subsection{Accounting for size-dependent surface effects}

Given that intracluster repulsions scale as expected (with \(N^{5/3}\)), the simplest extensive free energy expression (resembling that of Groenewold and Kegel) that readily %naturally 
leads to the empirically-observed scaling in Fig. 3 is one where the surface-energy penalty, rather than scaling as \(N^{2/3}\), instead effectively scales with a lesser exponent:

\begin{equation}~\label{eqn:reqscaling}
\beta \Delta F(N) = -N \beta \Delta \mu_{0}^{\text{SA}} + \nu_{1}\beta\varepsilon N^{1/3} + \dfrac{\beta A_{\text{MAX}}N^{5/3}}{2\alpha} 
\end{equation}

\noindent Here, \(\nu_{1}\) is some (as yet undetermined) dimensionless prefactor distinct from the \(\nu_{0}\) above. In turn, it is easily shown that solving Eqn.~\ref{eqn:reqscaling} for \(\beta\Delta f(N^{*}) \equiv \min_{N} [\beta\Delta f(N)]\) results in the generalized scaling \(N^{*} \propto \{\beta\varepsilon/[\beta A_{\text{MAX}}]\}^{3/4}\) or, in the unscreened limit, \(N^{*}_{\infty} \propto \{\beta\varepsilon/[Z^{2}(\lambda_{\text{B}}/d)]\}^{3/4}\).

During the remainder of this section, our ultimate goal is to demonstrate that this reduced exponent for the surface energy term naturally emerges for our clustered systems because the effective energy penalty is dependent on cluster size \(N^{*}\) in the range \(6 \leq N^{*} \leq 60\). 
% *** NEW
Conceptually, this size-dependence for the surface energy echoes the long-established notion that the generalized surface-tension of a liquid droplet with high curvature \(\gamma(R)\) will depart from the reference surface tension \(\gamma^{\infty}\) of a planar liquid-vapor interface (or very large droplet with low curvature). Indeed, starting with pioneering work by Tolman~\cite{Tolman1949}, a vast number of studies have been dedicated to measuring first- and/or second-order corrections for \(\gamma(R)/\gamma^{\infty}\) (the classic first order correction depends on the ``Tolman length'') to better model, e.g., homogeneous nucleation, but this topic continues to be active and challenging area of research even for model systems like the LJ fluid~\cite{Nijmeijer1992, McGraw1996, tenWolde1998, Koga1998, vanGiessen2009, Troster2012, Wilhelmsen2015}. Compared to these studies, which are especially difficult given their general focus on \emph{critically-unstable} droplet formation (usually droplets with radius \(R \approx 4d\) at the smallest), the following analysis is notable because we consider \emph{stable} droplets with effective surface tensions dominated by \emph{short-range} attractive bonds (much shorter than, e.g., LJ attraction range) and radii of less than three particle diameters.

%
% used to be connected to above paragraph
Specifically, to capture this size-dependent surface energy, one ought to account for an \(N\)-dependent number of missing coordination bonds \(z_{\text{c,m}}(N)\) for the surface particles relative to the reference bulk (interior) coordination number \(z_{\text{c,0}}\). The surface energy penalty in Eqn.~\ref{eqn:reqscaling} can then be written

\begin{equation}~\label{eqn:reqFsurf}
\nu_{1}\beta\varepsilon N^{1/3} \propto z_{\text{c,m}}(N) \beta\varepsilon N^{2/3}
\end{equation}

\noindent with the (to be demonstrated) scaling

\begin{equation}~\label{eqn:reqFsurfII}
z_{\text{c,m}}(N) \propto N^{-1/3}
\end{equation}

\noindent where we still assume that the number of surface particles at least roughly scales as \(N^{2/3}\), i.e., proportional to the squared cluster radius \((R_{\text{G}}/d)^{2} = \alpha^{2} N^{2/3}\), though making a formal distinction between interior and surface particles is difficult for small \(N\) (as discussed later). To demonstrate that the scaling in Eqn.~\ref{eqn:reqFsurfII} is reasonable, we show in Figs. 5 and 6 that this size-dependence for \(z_{\text{c,m}}(N) = z_{\text{c,0}} - z_{\text{c}}(N)\) originates based on the coordination number of (surface) particles \(z_{\text{c}}(N)\) measured from MD configurations, which we calculate from the extensive number of intracluster bonds \(n_{\text{B}}(N)\). Given our measurement of \(n_{\text{B}}(N)\) is at the root of much of this analysis, we consider its behavior first and proceed backwards to the scaling of Eqn.~\ref{eqn:reqFsurfII}.

Looking towards estimating \(z_{\text{c,m}}(N)\), consider in Fig. 5(a) the extensive number of intracluster bonds \(n_{\text{B}}(N)\) measured from MD simulations, where we observe a previously undiscovered (to our knowledge) superlinear growth rate over the range of cluster sizes that we generate. Interestingly, this superlinear behavior contrasts with known small- and large-cluster limits, which are linear in \(N\). Here, \(n_{\text{B}}(N)\) is nicely captured at each packing fraction for \(6 \leq N^{*} \leq 60\) by the empirical expression:

\begin{equation}~\label{eqn:nBgrowth}
n_{\text{B}}(N) = (k/2)N\ln(N)
\end{equation}

\noindent where \(k\) is a \(\phi\)-specific \(O(1)\) prefactor~\footnote{Note that the prefactor \(k\) modestly decreases as \(\phi\) increases: this occurs because, as discussed earlier, the cluster radius modestly increases with \(\phi\) for fixed \(N^{*}\); thus, clusters become less dense and exhibit correspondingly fewer bonds.} 
and we include a division by 2 for aesthetic alignment with the next results. This superlinear regime contrasts with the small cluster regime (\(3 \leq N \leq 9\)), where it is known~\cite{Arkus2009,Meng2010} that colloidal clusters dominated by SA bonds maximize their extensive bonding number according to the expression \(n_{\text{B}}(N) = 3N-6\). Likewise, in the limit of large droplets, the number of bonds must scale increasingly like in the corresponding bulk fluid, i.e., \(n_{\text{B}}(N) \rightarrow (z_{\text{bulk}}/2)N\) where \(z_{\text{bulk}}\) is the coordination number of the reference fluid (or crystal) phase.

\begin{figure}
\begin{center}
  \includegraphics[trim={0 0 0 0},clip]{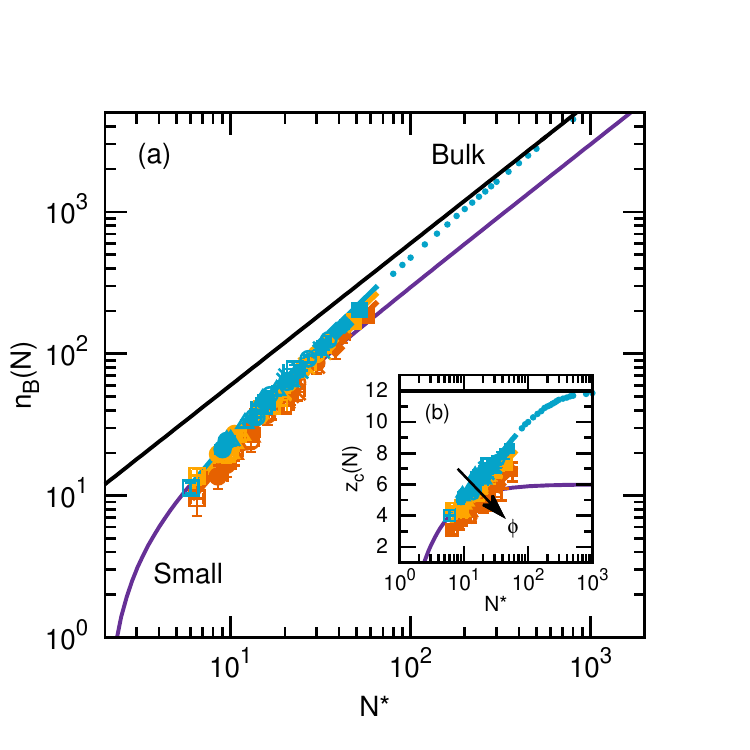}
  \caption{(a) Extensive number of intracluster bonds \(n_{\text{B}}(N)\) versus cluster size \(N^{*}\). Blue, yellow, and orange symbols correspond to measurements from simulations at \(\phi = 0.015\), 0.030, and 0.060, respectively. Symbol types correspond to constant charge \(Z\) as listed in Table I (note that we test various screening lengths \(\kappa^{-1}/d\) at each \(Z\)). Blue, yellow, and orange solid lines are of the empirical form \(n_{\text{B}}(N) = (k/2)N\ln(N)\) found to apply between \(6 \leq N^{*} \leq 60\), where \(k=2.20\), 1.95, and 1.70 with respect to \(\phi\). Purple line corresponds to small cluster limit~\cite{Arkus2009,Meng2010} \(n_{\text{B}}(N) = 3N-6\), which is accurate for \(3 \leq N \leq 9\). Black line corresponds to large droplet (bulk) limit \(n_{\text{B}}(N) = (z_{\text{bulk}}/2)N\) where we choose \(z_{\text{bulk}}=12\) (see text); this limit becomes near-quantitative for dense droplets of \(N \approx O(1000)\). Dashed blue curve is a schematic extension to the solid blue line between \(60 \leq N^{*} \leq 500\). (b) Average coordination number \(z_{\text{c}}(N)\) versus cluster size \(N^{*}\). Symbols and lines have same meaning as in (a), where the latter are calculated via the formula \(z_{\text{c}}(N) = 2n_{\text{B}}(N)/N\).}
  \label{sch:Figure-inftyscaling}
\end{center}
\end{figure}

To quickly understand why \(n_{\text{B}}(N)\) growth should be superlinear over this size range, we show in Fig 5(a) extensions of the small- and large-cluster linear regimes (to large and small \(N\) where they should respectively fail) to demonstrate that the function \(n_{\text{B}}(N) = (k/2)N\ln(N)\) connects these otherwise disparate limits while quantitatively overlapping with the upper reaches of the small cluster trend at \(N \approx 10\). To wit, notice that the characteristic slope of the small-\(N\) regime is \(m=3\) differs meaningfully from the typical slope in the large-\(N\) regime of a very dense bulk fluid or crystal, which we estimate as \(m = z_{\text{bulk}}/2 = 6\) with \(z_{\text{bulk}}=12\) because it is the sphere kissing number in three dimensions~\cite{Pfender2004} (this is justified later). Thus, provided \(z_{\text{bulk}}\) is decidedly larger than 3, a superlinear regime allows for a smooth continuous growth in \(n_{\text{B}}(N)\) with respect to \(N\).

This connectivity between very small and large cluster sizes is clearly echoed by the next necessary quantity we must calculate: the average coordination number \(z_{\text{c}}(N) = 2n_{\text{B}}(N)/N = k\ln(N)\), 
% *** NEW
which we show in Fig. 5(b) for all of our clustered states~\footnote{
The relation between coordination number and cluster size that we observe, \(z_{\text{c}}(N) = k\ln(N)\) (with \(k \approx 2\)), has a much stronger scaling than that of a similar relation reported by Godfrin et. al.~\cite{GodfrinWagnerLiu2014}, which was given as \(z_{\text{c}}(N) = 1.5[\ln(N)]^{1/2}\) (here written in our choice of notation). We would simply note that the latter reaches an \emph{asymptotic} coordination number of approximately 4 at very large droplet sizes, which would point to extremely elongated non-compact clusters (even Bernal spiral motifs~\cite{Campbell2005} exhibit \(z_{\text{c}} \approx 5\)). In contrast, our expression, which is based on data from compact spherical aggregates at the onset of clustering, grows with cluster size and tends to approach the bulk coordination number \(z_{\text{bulk}}=12\) of a dense attractive fluid in the large \(N\) limit, as in Fig. 5(b).}.
Here, we plot \(z_{\text{c}}(N)\) values calculated from MD configurations, which begin to bridge the gap (up to the highest cluster sizes we observe) between the highly bond-restricted regime at small \(N\) and the bulk regime at large \(N\) where the coordination number approaches \(z_{\text{c}}(N) \rightarrow z_{\text{bulk}}\). Notably, \(z_{\text{c}}(N)\) varies by approximately a \emph{factor of 2} over the size range of interest \(6 \leq N^{*} \leq 60\), which underlines that the conventional practice (for larger droplets) of assuming that surface effects are size-independent is problematic for these smaller aggregates.

With \(z_{\text{c}}(N)\) in hand, we can proceed to calculate the average number of missing bonds per particle \(z_{\text{c,m}}(N)\), which indeed collapses onto a master curve scaling as \(N^{-1/3}\) (shown in Fig. 6) when the magnitude of the reference (fitting) coordination parameter \(z_{\text{c,0}}\) is set--in line with measurements of cluster interiors--at values appropriate for highly-packed bulk fluids. To do this, we use the expression

\begin{equation}~\label{eqn:zcm}
z_{\text{c,m}}(N) = z_{\text{c,0}} - z_{\text{c}}(N) %\approx z_{\text{c,0}} - k\ln(N)
\end{equation}

\noindent where the only as-yet undetermined value is \(z_{\text{c,0}}\), which is the coordination number of the reference bulk SA fluid that represents the idealized cluster interior; for our immediate purposes, we treat this parameter as tunable and verify our choices as reasonable below. As shown in Fig. 6(a), our data approximately collapse onto a master curve with characteristic \(N^{-1/3}\) dependence when \(z_{\text{c,0}} = 12.0\), 11.5, and 10.5 for \(\phi = 0.015\), 0.030, and 0.060, respectively. All of these values--especially for the lowest-density case--are reflective of bulk fluids dominated by short-range attractions, especially here given that energetic gains from bonding occur within attractive wells beyond surface contact that are approximately \(0.1d\) in width.

\begin{figure}
\begin{center}
  \includegraphics[trim={0 0 0 0},clip]{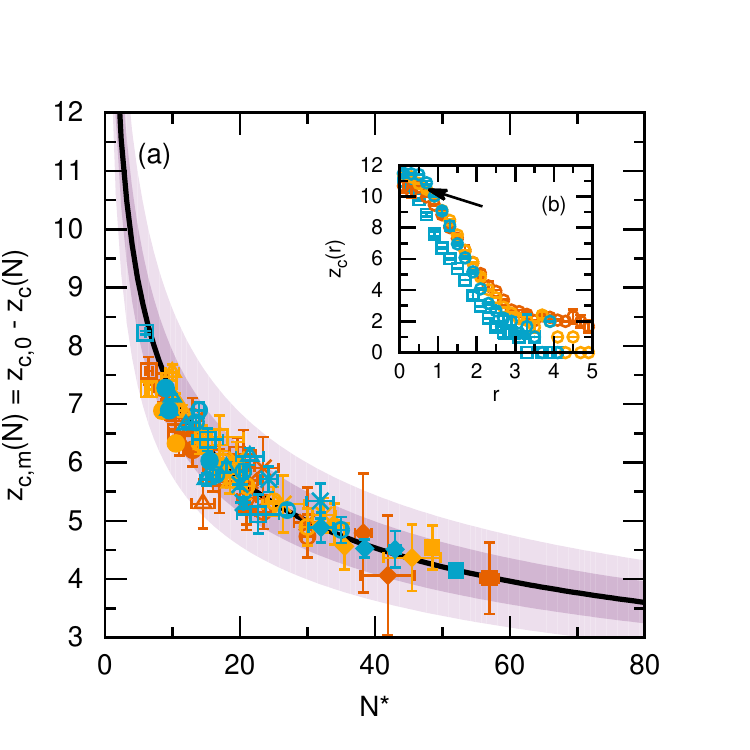}
  \caption{(a) Average number of missing bonds per particle \(z_{\text{c,m}}(N) = z_{\text{c,0}} - z_{\text{c}}(N)\) versus cluster size \(N^{*}\). Blue, yellow, and orange symbols correspond to measurements from simulations at \(\phi = 0.015\), 0.030, and 0.060, respectively. Fitting parameter \(z_{\text{c,0}}\) is the coordination number of the reference bulk dense fluid, found to be \(z_{\text{c,0}} = 12.0\), 11.5, and 10.5 with respect to \(\phi\). Thick black line is a scaling guideline with the form \(z_{\text{c,m}}(N) = 15.5 N^{-1/3}\) and dark (light) purple shadings correspond to 10\% (20\%) deviation from this scaling. Symbol types correspond to constant charge \(Z\) as listed in Table I (note that we test various screening lengths \(\kappa^{-1}/d\) at each \(Z\)). (b) Locally-averaged intracluster coordination number \(z_{\text{c}}(r)\) measured at radial positions \(r\) relative to cluster center of mass for four selected cluster phases. Blue, yellow, and orange circles are for \(Z=3.0\) and \(\kappa^{-1}/d=4.0\) at \(\phi = 0.015\), 0.030, and 0.060, respectively, where \(50 < N^{*} < 60\). Blue squares are for \(Z=6.0\) and \(\kappa^{-1}/d=4.0\) at \(\phi = 0.015\), where \(N^{*} \approx 20\). Arrow points to inner regions of clusters, highlighting \(z_{\text{c}}(r\rightarrow0) \approx 12\).}
  \label{sch:Figure-inftyscaling}
\end{center}
\end{figure}

In Fig. 6(b), we demonstrate that these \(z_{\text{c,0}}\) values are appropriate based on direct measurements of the locally-averaged coordination number \(z_{\text{c}}(r)\) as a function of radial position within clusters (relative to cluster center-of-mass). Here, we specifically show results from some of the largest clusters observed (\(50 < N^{*} < 60\)), which are most likely to possess bulk-like interiors as \(r \rightarrow 0\); indeed, it is evident that \(z_{\text{c}}(r\rightarrow0) \approx 12\) for these larger clusters, though the limiting value (as above) slightly decreases as \(\phi\) increases, presumably due to the previously-discussed trend in intracluster density. We also observe in Fig. 6(b) that the \(z_{\text{c}}(r\rightarrow0)\) limit is similar even for smaller clusters, e.g., \(N^{*} \approx 20\), where central particles can still be surrounded by a packed shell of intracluster neighbors.

Taken altogether, the results of Figs. 5 and 6 nicely justify the choice to quantify the surface energy penalty of cluster formation from the perspective of a size-dependence in the relative number of missing bonds \(z_{\text{c,m}}(N)\). Before moving on to consider the impact of the scaling relationships in Eqns.~\ref{eqn:reqFsurf} and \ref{eqn:reqFsurfII} for predicting cluster size, we pause to note that in the analysis above, we approximate \(z_{\text{c,m}}(N)\) for surface particles based on an average measurement of \(n_{\text{B}}(N)\) for \emph{all} cluster constituents. We take this somewhat imprecise approach because it draws upon relatively unambiguous measurable quantities and bypasses the fraught process of definitively distinguishing between surface and interior particles (consider, e.g., Fig. 6(b)). Our approximation is sufficient for the proof-of-concept analysis here, but we imagine a more exacting analysis in this vein would be a worthwhile future endeavor.~\footnote{In our approximate treatment, we suspect that \emph{at small} \(N\), we are simultaneously: (1) underestimating the relative fraction of surface particles, which means the number of ``surface'' particles actually scales as \(N^{m}\) with \(m<2/3\) over the whole intermediate size range; and (2) overestimating the coordination number \(z_{\text{c}}(N)\) of surface particles (underestimating \(z_{\text{c,m}}(N)\)), which means that the number of missing surface bonds actually scales as \(z_{\text{c,m}}(N) \propto N^{m}\) with \(m>-1/3\). Because these errors in the exponents tend to cancel each other, we expect that the net effective \(N^{1/3}\) scaling of the surface term in Eqn.~\ref{eqn:reqFsurf} holds even given greater precision in the configurational analysis.}

%%%%%%%%%%%%%%%
%%%%%%%%%%%%%%%
%%%%%%%%%%%%%%%
\subsection{Revised free energy model for predicting size}

Based on the new scaling for the free energy surface penalty justified in Section~\ref{sect:surf} and the generalized free energy term for repulsive contributions in Section~\ref{sect:rep}, we can return to the extensive free energy model of Eqn.~\ref{eqn:reqscaling} and readily derive a new master equation for predicting cluster size \(N^{*}\) based on experimentally-tunable parameters:

\begin{equation}~\label{eqn:finalpredictivemine}
N^{*} = \Bigg[\dfrac{\alpha\nu_{2}\beta\varepsilon}{\beta A_{\text{MAX}}}\Bigg]^{3/4} = \Bigg[\dfrac{\alpha\nu_{2}\beta\varepsilon\{1.0+0.5/(\kappa^{-1}/d)\}^2}{Z^{2}(\lambda_{\text{B}}/d)}\Bigg]^{3/4}
\end{equation}

\noindent where, as before, \(\alpha\) is the known \(\phi\)-dependent prefactor relating cluster radius and number size and \(\nu_{2}\) is a constant similar to those above that scales the surface energy penalty, which we treat as an empirical tuning parameter. Eqn.~\ref{eqn:finalpredictivemine} is the central result of this publication.

\begin{figure}
\begin{center}
  \includegraphics[resolution=300,trim={0 0 0 0},clip]{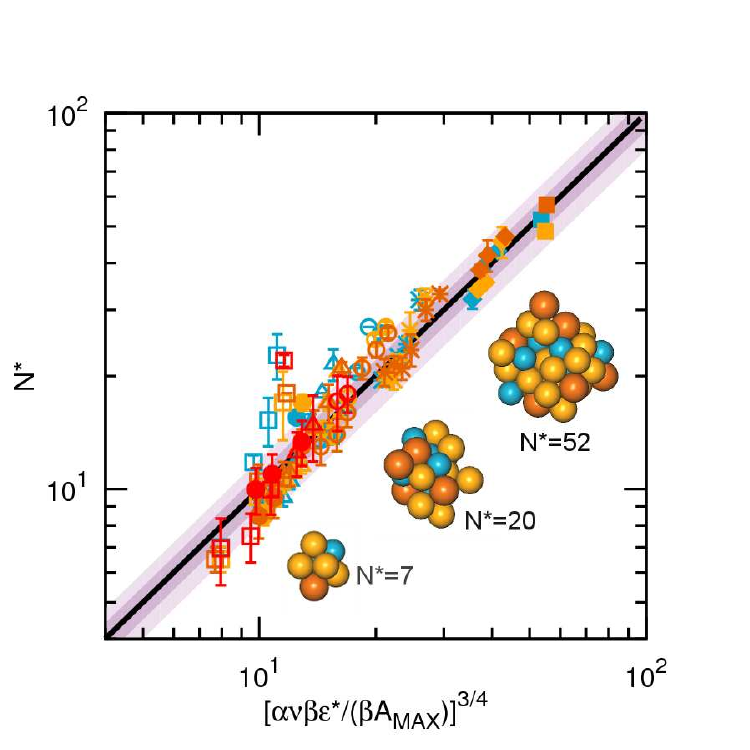}
  \caption{Measured cluster size \(N^{*}\) versus predicted cluster size from Eqn.~\ref{eqn:finalpredictivemine}, where the latter formula is a function of \(\phi\)-dependent radius of gyration coefficient \(\alpha\) (see Fig. 2); critical attraction strength \(\beta\varepsilon^{*}\); and maximum repulsive barrier height \(\beta A_{\text{MAX}} = Z^{2}(\lambda_{\text{B}}/d)/[1.0+0.5/(\kappa^{-1}/d)]^2\). The constant prefactor \(\nu_{2} \simeq \sqrt{2}\pi\) scales the surface energy penalty associated with aggregation (see text). Thick black line corresponds to the Eqn.~\ref{eqn:finalpredictivemine} relation and dark (light) purple shadings correspond to 10\% (20\%) deviation from this scaling. Blue, yellow, orange, and red symbols correspond to results from simulations at \(\phi = 0.015\), 0.030, 0.060, and 0.120, respectively. Symbol types correspond to constant charge \(Z\) as listed in Table I (note that we test various screening lengths \(\kappa^{-1}/d\) at each \(Z\)). The three illustrated clusters are instantaneous configurations observed in MD simulations, with blue, yellow, and orange spheres corresponding to small, medium, and large particles, respectively.}
  \label{sch:Figure-inftyscaling}
\end{center}
\end{figure}

As demonstrated in Fig. 7, Eqn.~\ref{eqn:finalpredictivemine} successfully predicts characteristic cluster sizes for the vast majority of our \({\approx}100\) cases over various \(Z\)-\(\kappa^{-1}/d\) combinations and near order-of-magnitude ranges in both size \(6 \leq N^{*} \leq 60\) and bulk monomer packing fraction \(0.015 \leq \phi \leq 0.120\). This wide applicability is notable as the underlying free energy framework remains very simple: to wit, intercluster effects can evidently be neglected even as conditions become less dilute (e.g., \(\phi \approx 0.120\)), though the current model cannot predict more subtle trends known for the SALR model~\cite{JadrichSM2015} like the growing polydispersity of aggregates with increasing \(\phi\). Meanwhile, the biggest deviations between measured and predicted \(N^{*}\) (larger than 20\%) occur for states combining large charge (e.g., \(Z=15.0\)) and small screening length (e.g., \(\kappa^{-1}/d = 0.70\)), which result in rather non-idealized repulsions that are both strong relative to \(k_{\text{B}}T\) and far from the Coulombic limit. Finally, note that the value for prefactor \(\nu_{2}\) that shifts the (already collapsed) predictions into the correct range is \(\nu_{2} \approx \sqrt{2}\pi\), which we expect should apply rather generally for compact colloidal clusters as it 
simply converts between measurements of the cluster surface-size based on population and radius.
%simply accounts for the reference areal footprint of each surface monomer relative to how the cluster surface area grows with \(N\).

% &
% &&&
% &&&&&
% ===============================================================================================================
\section{Conclusions}
% ===============================================================================================================
% &&&&&
% &&&
% &

We have validated a new and readily applied formula (Eqn.~\ref{eqn:finalpredictivemine}) that can predict characteristic cluster size \(N^{*}\) for idealized SALR suspensions as a function of the variables controlling monomer-monomer interactions (including attraction strength \(\beta\varepsilon\), surface charge \(Z\), and screening length \(\kappa^{-1}/d\)). Eqn.~\ref{eqn:finalpredictivemine} and its underlying free energy model represent a semi-empirical adaptation and extension of the canonical free energy model due to Groenewold and Kegel~\cite{GroenewoldKegel2001,Groenewold2004,Zhang2012}, where we found the latter exhibits a spurious scaling of \(N^{*}\) away from the large-droplet limit with respect to the ratio of attractive and repulsive interaction strengths driving aggregation. We subsequently find that Eqn.~\ref{eqn:finalpredictivemine} performs excellently based on direct comparisons of predicted cluster sizes and measurements of \(N^{*}\) from MD simulations of approximately 100 different systems for very wide ranges in \(\phi\), \(Z\), and \(\kappa^{-1}/d\), where we examine states at the onset of clustering that exhibit compact spherical aggregates in the size range \(6 \leq N^{*} \leq 60\).

The predictive quality of Eqn.~\ref{eqn:finalpredictivemine} demonstrates that a simple free energy model, which treats SALR systems as reference SA fluids (via classical nucleation theory) with additive repulsive perturbations due to electrostatic effects, can be applied down to \emph{extremely small} cluster sizes (\(N^{*} < 10\)) provided one properly corrects for surface effects at small \(N^{*}\). Conceptually, this is in the spirit of long-standing investigations regarding size-dependent surface tensions in small droplets~\cite{Tolman1949,Nijmeijer1992, McGraw1996, tenWolde1998, Koga1998, vanGiessen2009, Troster2012, Wilhelmsen2015}, and practically, we find that one can treat the energy penalty of ``interface'' formation as a function of an \(N\)-dependent number of missing coordination bonds \(z_{\text{c,m}}(N)\) for surface particles (referenced against the coordination number in the bulk fluid). Here, this picture is validated in part by configurational analysis of the number of extensive intracluster bonds \(n_{\text{B}}(N)\), which revealed a previous undiscovered (to our knowledge) \emph{superlinear} scaling regime for \(n_{\text{B}}(N)\) over the size range \(6 \leq N^{*} \leq 60\). Meanwhile, based on the form of the free energy model, we confirm that intercluster effects can be neglected even for rather non-dilute conditions (e.g., \(\phi = 0.120\)), which is reflected by our observation that cluster size \(N^{*}\) exhibits little variability with respect to \(\phi\) given otherwise fixed conditions.

We look forward to testing the predictive capability of Eqn.~\ref{eqn:finalpredictivemine} for real colloidal suspensions that exhibit equilibrium cluster phases, which could help bolster whether SALR pair potentials are a sufficient (if idealized) description of experimental systems. For instance, there has been recent discussion~\cite{ParkGlotzer2012,Sweatman2014,NguyenGlotzer2015} in the literature as to whether accounting for charge renormalization during aggregation is necessary for describing cluster behavior; likewise, Groenewold and Kegel initially postulated that non-trivial charge effects~\cite{Manning1979,Alexander1984,Ramanathan1988,Gillespie2014} could affect the free energy picture in certain limits. Of course, these effects are not captured by the canonical SALR pair potential examined here, but we now possess a free energy model (Eqn.~\ref{eqn:reqscaling}) known to describe this simpler system. Thus, ascertaining whether cluster size \(N^{*}\) scales in experiments similarly to the empirical scaling of Eqn.~\ref{eqn:finalpredictivemine} would help clarify the degree to which phenomenology and interpretive guidelines derived from the pairwise model are appropriate for real systems. Similarly, it is fascinating to consider how accounting for size-dependent surface effects, here so crucial for producing quantitative predictions, might change for less compact (e.g., elongated) aggregates than those considered here.

% &
% &&&
% &&&&&
% ===============================================================================================================
\section{Acknowledgments}
% ===============================================================================================================
% &&&&&
% &&&
% &

This work was partially supported by the National Science Foundation (1247945) and the Welch Foundation (F-1696). We acknowledge the Texas Advanced Computing Center (TACC) at The University of Texas at Austin for providing HPC resources.

%\bibliography{Bibliography}

\begin{thebibliography}{62}%
\makeatletter
\providecommand \@ifxundefined [1]{%
 \@ifx{#1\undefined}
}%
\providecommand \@ifnum [1]{%
 \ifnum #1\expandafter \@firstoftwo
 \else \expandafter \@secondoftwo
 \fi
}%
\providecommand \@ifx [1]{%
 \ifx #1\expandafter \@firstoftwo
 \else \expandafter \@secondoftwo
 \fi
}%
\providecommand \natexlab [1]{#1}%
\providecommand \enquote  [1]{``#1''}%
\providecommand \bibnamefont  [1]{#1}%
\providecommand \bibfnamefont [1]{#1}%
\providecommand \citenamefont [1]{#1}%
\providecommand \href@noop [0]{\@secondoftwo}%
\providecommand \href [0]{\begingroup \@sanitize@url \@href}%
\providecommand \@href[1]{\@@startlink{#1}\@@href}%
\providecommand \@@href[1]{\endgroup#1\@@endlink}%
\providecommand \@sanitize@url [0]{\catcode `\\12\catcode `\$12\catcode
  `\&12\catcode `\#12\catcode `\^12\catcode `\_12\catcode `\%12\relax}%
\providecommand \@@startlink[1]{}%
\providecommand \@@endlink[0]{}%
\providecommand \url  [0]{\begingroup\@sanitize@url \@url }%
\providecommand \@url [1]{\endgroup\@href {#1}{\urlprefix }}%
\providecommand \urlprefix  [0]{URL }%
\providecommand \Eprint [0]{\href }%
\providecommand \doibase [0]{http://dx.doi.org/}%
\providecommand \selectlanguage [0]{\@gobble}%
\providecommand \bibinfo  [0]{\@secondoftwo}%
\providecommand \bibfield  [0]{\@secondoftwo}%
\providecommand \translation [1]{[#1]}%
\providecommand \BibitemOpen [0]{}%
\providecommand \bibitemStop [0]{}%
\providecommand \bibitemNoStop [0]{.\EOS\space}%
\providecommand \EOS [0]{\spacefactor3000\relax}%
\providecommand \BibitemShut  [1]{\csname bibitem#1\endcsname}%
\let\auto@bib@innerbib\@empty
%</preamble>
\bibitem [{\citenamefont {Smoluchowski}(1916)}]{Smoluchowski1916}%
  \BibitemOpen
  \bibfield  {author} {\bibinfo {author} {\bibfnamefont {M.~V.}\ \bibnamefont
  {Smoluchowski}},\ }\bibfield  {title} {\enquote {\bibinfo {title} {Drei
  vortr\"{a}ge \"{u}ber diffusion, brownsche bewegung und koagulation von
  kolloidteilchen},}\ }\href@noop {} {\bibfield  {journal} {\bibinfo  {journal}
  {Physik Z.}\ }\textbf {\bibinfo {volume} {17}},\ \bibinfo {pages} {557--585}
  (\bibinfo {year} {1916})}\BibitemShut {NoStop}%
\bibitem [{\citenamefont {Derjaguin}\ and\ \citenamefont
  {Landau}(1941)}]{DerjaguinLandau1941}%
  \BibitemOpen
  \bibfield  {author} {\bibinfo {author} {\bibfnamefont {B.~V.}\ \bibnamefont
  {Derjaguin}}\ and\ \bibinfo {author} {\bibfnamefont {L.}~\bibnamefont
  {Landau}},\ }\bibfield  {title} {\enquote {\bibinfo {title} {Theory of the
  stability of strongly charged lyophobic sols and of the adhesion of strongly
  charged particles in solution of electrolytes},}\ }\href@noop {} {\bibfield
  {journal} {\bibinfo  {journal} {Acta Physicochim. URSS}\ }\textbf {\bibinfo
  {volume} {14}},\ \bibinfo {pages} {633--662} (\bibinfo {year}
  {1941})}\BibitemShut {NoStop}%
\bibitem [{\citenamefont {Verwey}\ and\ \citenamefont
  {Overbeek}(1948)}]{VerweyOverbeek1948}%
  \BibitemOpen
  \bibfield  {author} {\bibinfo {author} {\bibfnamefont {E.~J.}\ \bibnamefont
  {Verwey}}\ and\ \bibinfo {author} {\bibfnamefont {J.~T.~G.}\ \bibnamefont
  {Overbeek}},\ }\href@noop {} {\emph {\bibinfo {title} {Theory of the
  Stability Lyophobic Colloids}}}\ (\bibinfo  {publisher} {Elsevier},\ \bibinfo
  {address} {New York, NY, USA},\ \bibinfo {year} {1948})\BibitemShut {NoStop}%
\bibitem [{\citenamefont {Lin}\ \emph {et~al.}(1989)\citenamefont {Lin},
  \citenamefont {Lindsay}, \citenamefont {A.}, \citenamefont {Ball},
  \citenamefont {Klein},\ and\ \citenamefont {Meakin}}]{Lin1989}%
  \BibitemOpen
  \bibfield  {author} {\bibinfo {author} {\bibfnamefont {M.~Y.}\ \bibnamefont
  {Lin}}, \bibinfo {author} {\bibfnamefont {H.~M.}\ \bibnamefont {Lindsay}},
  \bibinfo {author} {\bibfnamefont {W.~D.}\ \bibnamefont {A.}}, \bibinfo
  {author} {\bibfnamefont {R.~C.}\ \bibnamefont {Ball}}, \bibinfo {author}
  {\bibfnamefont {R.}~\bibnamefont {Klein}}, \ and\ \bibinfo {author}
  {\bibfnamefont {P.}~\bibnamefont {Meakin}},\ }\bibfield  {title} {\enquote
  {\bibinfo {title} {Universality in colloid aggregation},}\ }\href {\doibase
  doi:10.1038/339360a0} {\bibfield  {journal} {\bibinfo  {journal} {Nature}\
  }\textbf {\bibinfo {volume} {339}},\ \bibinfo {pages} {360--362} (\bibinfo
  {year} {1989})}\BibitemShut {NoStop}%
\bibitem [{\citenamefont {Anderson}\ and\ \citenamefont
  {Lekkerkerker}(2002)}]{Anderson2002}%
  \BibitemOpen
  \bibfield  {author} {\bibinfo {author} {\bibfnamefont {V.}~\bibnamefont
  {Anderson}}\ and\ \bibinfo {author} {\bibfnamefont {H.~N.~W.}\ \bibnamefont
  {Lekkerkerker}},\ }\bibfield  {title} {\enquote {\bibinfo {title} {Insights
  into phase transition kinetics from colloid science},}\ }\href {\doibase
  doi:10.1038/416811a} {\bibfield  {journal} {\bibinfo  {journal} {Nature}\
  }\textbf {\bibinfo {volume} {416}},\ \bibinfo {pages} {811--815} (\bibinfo
  {year} {2002})}\BibitemShut {NoStop}%
\bibitem [{\citenamefont {Israelachvili}(2011)}]{Israelachvili2011}%
  \BibitemOpen
  \bibfield  {author} {\bibinfo {author} {\bibfnamefont {J.~N.}\ \bibnamefont
  {Israelachvili}},\ }\href@noop {} {\emph {\bibinfo {title} {Intermolecular
  and Surface Forces}}}\ (\bibinfo  {publisher} {Academic Press},\ \bibinfo
  {address} {New York, NY, USA},\ \bibinfo {year} {2011})\BibitemShut {NoStop}%
\bibitem [{\citenamefont {Groenewold}\ and\ \citenamefont
  {Kegel}(2001)}]{GroenewoldKegel2001}%
  \BibitemOpen
  \bibfield  {author} {\bibinfo {author} {\bibfnamefont {J.}~\bibnamefont
  {Groenewold}}\ and\ \bibinfo {author} {\bibfnamefont {W.~K.}\ \bibnamefont
  {Kegel}},\ }\bibfield  {title} {\enquote {\bibinfo {title} {Anomalously large
  equilibrium clusters of colloids†},}\ }\href {\doibase 10.1021/jp011646w}
  {\bibfield  {journal} {\bibinfo  {journal} {J. Phys. Chem. B}\ }\textbf
  {\bibinfo {volume} {105}},\ \bibinfo {pages} {11702--11709} (\bibinfo {year}
  {2001})},\ \Eprint {http://arxiv.org/abs/http://dx.doi.org/10.1021/jp011646w}
  {http://dx.doi.org/10.1021/jp011646w} \BibitemShut {NoStop}%
\bibitem [{\citenamefont {Sciortino}\ \emph {et~al.}(2004)\citenamefont
  {Sciortino}, \citenamefont {Mossa}, \citenamefont {Zaccarelli},\ and\
  \citenamefont {Tartaglia}}]{Sciortino2004}%
  \BibitemOpen
  \bibfield  {author} {\bibinfo {author} {\bibfnamefont {F.}~\bibnamefont
  {Sciortino}}, \bibinfo {author} {\bibfnamefont {S.}~\bibnamefont {Mossa}},
  \bibinfo {author} {\bibfnamefont {E.}~\bibnamefont {Zaccarelli}}, \ and\
  \bibinfo {author} {\bibfnamefont {P.}~\bibnamefont {Tartaglia}},\ }\bibfield
  {title} {\enquote {\bibinfo {title} {Equilibrium cluster phases and
  low-density arrested disordered states: The role of short-range attraction
  and long-range repulsion},}\ }\href {\doibase 10.1103/PhysRevLett.93.055701}
  {\bibfield  {journal} {\bibinfo  {journal} {Phys. Rev. Lett.}\ }\textbf
  {\bibinfo {volume} {93}},\ \bibinfo {pages} {055701} (\bibinfo {year}
  {2004})}\BibitemShut {NoStop}%
\bibitem [{\citenamefont {Archer}\ and\ \citenamefont
  {Wilding}(2007)}]{ArcherWilding2007}%
  \BibitemOpen
  \bibfield  {author} {\bibinfo {author} {\bibfnamefont {A.~J.}\ \bibnamefont
  {Archer}}\ and\ \bibinfo {author} {\bibfnamefont {N.~B.}\ \bibnamefont
  {Wilding}},\ }\bibfield  {title} {\enquote {\bibinfo {title} {Phase behavior
  of a fluid with competing attractive and repulsive interactions},}\ }\href
  {\doibase 10.1103/PhysRevE.76.031501} {\bibfield  {journal} {\bibinfo
  {journal} {Phys. Rev. E}\ }\textbf {\bibinfo {volume} {76}},\ \bibinfo
  {pages} {031501} (\bibinfo {year} {2007})}\BibitemShut {NoStop}%
\bibitem [{\citenamefont {Toledano}, \citenamefont {Sciortino},\ and\
  \citenamefont {Zaccarelli}(2009)}]{ToledanoSciortino2009}%
  \BibitemOpen
  \bibfield  {author} {\bibinfo {author} {\bibfnamefont {J.~C.~F.}\
  \bibnamefont {Toledano}}, \bibinfo {author} {\bibfnamefont {F.}~\bibnamefont
  {Sciortino}}, \ and\ \bibinfo {author} {\bibfnamefont {E.}~\bibnamefont
  {Zaccarelli}},\ }\bibfield  {title} {\enquote {\bibinfo {title} {Colloidal
  systems with competing interactions: from an arrested repulsive cluster phase
  to a gel},}\ }\href {\doibase 10.1039/B818169A} {\bibfield  {journal}
  {\bibinfo  {journal} {Soft Matter}\ }\textbf {\bibinfo {volume} {5}},\
  \bibinfo {pages} {2390--2398} (\bibinfo {year} {2009})}\BibitemShut {NoStop}%
\bibitem [{\citenamefont {Jiang}\ and\ \citenamefont {Wu}(2009)}]{JiangWu2009}%
  \BibitemOpen
  \bibfield  {author} {\bibinfo {author} {\bibfnamefont {T.}~\bibnamefont
  {Jiang}}\ and\ \bibinfo {author} {\bibfnamefont {J.}~\bibnamefont {Wu}},\
  }\bibfield  {title} {\enquote {\bibinfo {title} {Cluster formation and bulk
  phase behavior of colloidal dispersions},}\ }\href {\doibase
  10.1103/PhysRevE.80.021401} {\bibfield  {journal} {\bibinfo  {journal} {Phys.
  Rev. E}\ }\textbf {\bibinfo {volume} {80}},\ \bibinfo {pages} {021401}
  (\bibinfo {year} {2009})}\BibitemShut {NoStop}%
\bibitem [{\citenamefont {Bomont}\ \emph {et~al.}(2012)\citenamefont {Bomont},
  \citenamefont {Bretonnet}, \citenamefont {Costa},\ and\ \citenamefont
  {Hansen}}]{Bomont2012}%
  \BibitemOpen
  \bibfield  {author} {\bibinfo {author} {\bibfnamefont {J.-M.}\ \bibnamefont
  {Bomont}}, \bibinfo {author} {\bibfnamefont {J.-L.}\ \bibnamefont
  {Bretonnet}}, \bibinfo {author} {\bibfnamefont {D.}~\bibnamefont {Costa}}, \
  and\ \bibinfo {author} {\bibfnamefont {J.-P.}\ \bibnamefont {Hansen}},\
  }\bibfield  {title} {\enquote {\bibinfo {title} {Communication: Thermodynamic
  signatures of cluster formation in fluids with competing interactions},}\
  }\href {\doibase http://dx.doi.org/10.1063/1.4733390} {\bibfield  {journal}
  {\bibinfo  {journal} {J. Chem. Phys.}\ }\textbf {\bibinfo {volume} {137}},\
  \bibinfo {eid} {011101} (\bibinfo {year} {2012})}\BibitemShut {NoStop}%
\bibitem [{\citenamefont {Godfrin}\ \emph {et~al.}(2013)\citenamefont
  {Godfrin}, \citenamefont {Castañeda-Priego}, \citenamefont {Liu},\ and\
  \citenamefont {Wagner}}]{Godfrin2013}%
  \BibitemOpen
  \bibfield  {author} {\bibinfo {author} {\bibfnamefont {P.~D.}\ \bibnamefont
  {Godfrin}}, \bibinfo {author} {\bibfnamefont {R.}~\bibnamefont
  {Castañeda-Priego}}, \bibinfo {author} {\bibfnamefont {Y.}~\bibnamefont
  {Liu}}, \ and\ \bibinfo {author} {\bibfnamefont {N.~J.}\ \bibnamefont
  {Wagner}},\ }\bibfield  {title} {\enquote {\bibinfo {title} {Intermediate
  range order and structure in colloidal dispersions with competing
  interactions},}\ }\href {\doibase http://dx.doi.org/10.1063/1.4824487}
  {\bibfield  {journal} {\bibinfo  {journal} {J. Chem. Phys.}\ }\textbf
  {\bibinfo {volume} {139}},\ \bibinfo {eid} {154904} (\bibinfo {year}
  {2013})}\BibitemShut {NoStop}%
\bibitem [{\citenamefont {Godfrin}\ \emph {et~al.}(2014)\citenamefont
  {Godfrin}, \citenamefont {Valadez-Perez}, \citenamefont {Castaneda-Priego},
  \citenamefont {Wagner},\ and\ \citenamefont {Liu}}]{GodfrinWagnerLiu2014}%
  \BibitemOpen
  \bibfield  {author} {\bibinfo {author} {\bibfnamefont {P.~D.}\ \bibnamefont
  {Godfrin}}, \bibinfo {author} {\bibfnamefont {N.~E.}\ \bibnamefont
  {Valadez-Perez}}, \bibinfo {author} {\bibfnamefont {R.}~\bibnamefont
  {Castaneda-Priego}}, \bibinfo {author} {\bibfnamefont {N.~J.}\ \bibnamefont
  {Wagner}}, \ and\ \bibinfo {author} {\bibfnamefont {Y.}~\bibnamefont {Liu}},\
  }\bibfield  {title} {\enquote {\bibinfo {title} {Generalized phase behavior
  of cluster formation in colloidal dispersions with competing interactions},}\
  }\href {\doibase 10.1039/C3SM53220H} {\bibfield  {journal} {\bibinfo
  {journal} {Soft Matter}\ }\textbf {\bibinfo {volume} {10}},\ \bibinfo {pages}
  {5061--5071} (\bibinfo {year} {2014})}\BibitemShut {NoStop}%
\bibitem [{\citenamefont {Mani}\ \emph {et~al.}(2014)\citenamefont {Mani},
  \citenamefont {Lechner}, \citenamefont {Kegel},\ and\ \citenamefont
  {Bolhuis}}]{ManiBolhuis2014}%
  \BibitemOpen
  \bibfield  {author} {\bibinfo {author} {\bibfnamefont {E.}~\bibnamefont
  {Mani}}, \bibinfo {author} {\bibfnamefont {W.}~\bibnamefont {Lechner}},
  \bibinfo {author} {\bibfnamefont {W.~K.}\ \bibnamefont {Kegel}}, \ and\
  \bibinfo {author} {\bibfnamefont {P.~G.}\ \bibnamefont {Bolhuis}},\
  }\bibfield  {title} {\enquote {\bibinfo {title} {Equilibrium and
  non-equilibrium cluster phases in colloids with competing interactions},}\
  }\href {\doibase 10.1039/C3SM53058B} {\bibfield  {journal} {\bibinfo
  {journal} {Soft Matter}\ }\textbf {\bibinfo {volume} {10}},\ \bibinfo {pages}
  {4479--4486} (\bibinfo {year} {2014})}\BibitemShut {NoStop}%
\bibitem [{\citenamefont {Sweatman}, \citenamefont {Fartaria},\ and\
  \citenamefont {Lue}(2014)}]{Sweatman2014}%
  \BibitemOpen
  \bibfield  {author} {\bibinfo {author} {\bibfnamefont {M.~B.}\ \bibnamefont
  {Sweatman}}, \bibinfo {author} {\bibfnamefont {R.}~\bibnamefont {Fartaria}},
  \ and\ \bibinfo {author} {\bibfnamefont {L.}~\bibnamefont {Lue}},\ }\bibfield
   {title} {\enquote {\bibinfo {title} {Cluster formation in fluids with
  competing short-range and long-range interactions},}\ }\href {\doibase
  http://dx.doi.org/10.1063/1.4869109} {\bibfield  {journal} {\bibinfo
  {journal} {J. Chem. Phys.}\ }\textbf {\bibinfo {volume} {140}},\ \bibinfo
  {eid} {124508} (\bibinfo {year} {2014})}\BibitemShut {NoStop}%
\bibitem [{\citenamefont {Jadrich}\ \emph
  {et~al.}(2015{\natexlab{a}})\citenamefont {Jadrich}, \citenamefont
  {Bollinger}, \citenamefont {Johnston},\ and\ \citenamefont
  {Truskett}}]{JadrichBollinger2015}%
  \BibitemOpen
  \bibfield  {author} {\bibinfo {author} {\bibfnamefont {R.~B.}\ \bibnamefont
  {Jadrich}}, \bibinfo {author} {\bibfnamefont {J.~A.}\ \bibnamefont
  {Bollinger}}, \bibinfo {author} {\bibfnamefont {K.~P.}\ \bibnamefont
  {Johnston}}, \ and\ \bibinfo {author} {\bibfnamefont {T.~M.}\ \bibnamefont
  {Truskett}},\ }\bibfield  {title} {\enquote {\bibinfo {title} {Origin and
  detection of microstructural clustering in fluids with spatial-range
  competitive interactions},}\ }\href {\doibase 10.1103/PhysRevE.91.042312}
  {\bibfield  {journal} {\bibinfo  {journal} {Phys. Rev. E}\ }\textbf {\bibinfo
  {volume} {91}},\ \bibinfo {pages} {042312} (\bibinfo {year}
  {2015}{\natexlab{a}})}\BibitemShut {NoStop}%
\bibitem [{\citenamefont {Nguyen}\ \emph {et~al.}(2015)\citenamefont {Nguyen},
  \citenamefont {Schultz}, \citenamefont {Kotov},\ and\ \citenamefont
  {Glotzer}}]{NguyenGlotzer2015}%
  \BibitemOpen
  \bibfield  {author} {\bibinfo {author} {\bibfnamefont {T.~D.}\ \bibnamefont
  {Nguyen}}, \bibinfo {author} {\bibfnamefont {B.~A.}\ \bibnamefont {Schultz}},
  \bibinfo {author} {\bibfnamefont {N.~A.}\ \bibnamefont {Kotov}}, \ and\
  \bibinfo {author} {\bibfnamefont {S.~C.}\ \bibnamefont {Glotzer}},\
  }\bibfield  {title} {\enquote {\bibinfo {title} {Generic, phenomenological,
  on-the-fly renormalized repulsion model for self-limited organization of
  terminal supraparticle assemblies},}\ }\href {\doibase
  10.1073/pnas.1509239112} {\bibfield  {journal} {\bibinfo  {journal} {Proc.
  Natl. Acad. Sci. U. S. A.}\ }\textbf {\bibinfo {volume} {112}},\ \bibinfo
  {pages} {E3161--E3168} (\bibinfo {year} {2015})},\ \Eprint
  {http://arxiv.org/abs/http://www.pnas.org/content/112/25/E3161.full.pdf}
  {http://www.pnas.org/content/112/25/E3161.full.pdf} \BibitemShut {NoStop}%
\bibitem [{\citenamefont {Zhuang}\ and\ \citenamefont
  {Charbonneau}(2016)}]{ZhuangCharbonneau2016}%
  \BibitemOpen
  \bibfield  {author} {\bibinfo {author} {\bibfnamefont {Y.}~\bibnamefont
  {Zhuang}}\ and\ \bibinfo {author} {\bibfnamefont {P.}~\bibnamefont
  {Charbonneau}},\ }\bibfield  {title} {\enquote {\bibinfo {title} {Recent
  advances in the theory and simulation of model colloidal microphase
  formers},}\ }\href@noop {} {\bibfield  {journal} {\bibinfo  {journal}
  {Pre-print}\ } (\bibinfo {year} {2016})},\ \Eprint
  {http://arxiv.org/abs/arXiv:1605.09718} {arXiv:1605.09718} \BibitemShut
  {NoStop}%
\bibitem [{\citenamefont {Campbell}\ \emph {et~al.}(2005)\citenamefont
  {Campbell}, \citenamefont {Anderson}, \citenamefont {van Duijneveldt},\ and\
  \citenamefont {Bartlett}}]{Campbell2005}%
  \BibitemOpen
  \bibfield  {author} {\bibinfo {author} {\bibfnamefont {A.~I.}\ \bibnamefont
  {Campbell}}, \bibinfo {author} {\bibfnamefont {V.~J.}\ \bibnamefont
  {Anderson}}, \bibinfo {author} {\bibfnamefont {J.~S.}\ \bibnamefont {van
  Duijneveldt}}, \ and\ \bibinfo {author} {\bibfnamefont {P.}~\bibnamefont
  {Bartlett}},\ }\bibfield  {title} {\enquote {\bibinfo {title} {Dynamical
  arrest in attractive colloids: The effect of long-range repulsion},}\ }\href
  {\doibase 10.1103/PhysRevLett.94.208301} {\bibfield  {journal} {\bibinfo
  {journal} {Phys. Rev. Lett.}\ }\textbf {\bibinfo {volume} {94}},\ \bibinfo
  {pages} {208301} (\bibinfo {year} {2005})}\BibitemShut {NoStop}%
\bibitem [{\citenamefont {Klix}, \citenamefont {Royall},\ and\ \citenamefont
  {Tanaka}(2010)}]{Klix2010}%
  \BibitemOpen
  \bibfield  {author} {\bibinfo {author} {\bibfnamefont {C.~L.}\ \bibnamefont
  {Klix}}, \bibinfo {author} {\bibfnamefont {C.~P.}\ \bibnamefont {Royall}}, \
  and\ \bibinfo {author} {\bibfnamefont {H.}~\bibnamefont {Tanaka}},\
  }\bibfield  {title} {\enquote {\bibinfo {title} {Structural and dynamical
  features of multiple metastable glassy states in a colloidal system with
  competing interactions},}\ }\href {\doibase 10.1103/PhysRevLett.104.165702}
  {\bibfield  {journal} {\bibinfo  {journal} {Phys. Rev. Lett.}\ }\textbf
  {\bibinfo {volume} {104}},\ \bibinfo {pages} {165702} (\bibinfo {year}
  {2010})}\BibitemShut {NoStop}%
\bibitem [{\citenamefont {Zhang}\ \emph {et~al.}(2012)\citenamefont {Zhang},
  \citenamefont {Klok}, \citenamefont {Hans~Tromp}, \citenamefont
  {Groenewold},\ and\ \citenamefont {Kegel}}]{Zhang2012}%
  \BibitemOpen
  \bibfield  {author} {\bibinfo {author} {\bibfnamefont {T.~H.}\ \bibnamefont
  {Zhang}}, \bibinfo {author} {\bibfnamefont {J.}~\bibnamefont {Klok}},
  \bibinfo {author} {\bibfnamefont {R.}~\bibnamefont {Hans~Tromp}}, \bibinfo
  {author} {\bibfnamefont {J.}~\bibnamefont {Groenewold}}, \ and\ \bibinfo
  {author} {\bibfnamefont {W.~K.}\ \bibnamefont {Kegel}},\ }\bibfield  {title}
  {\enquote {\bibinfo {title} {Non-equilibrium cluster states in colloids with
  competing interactions},}\ }\href {\doibase 10.1039/C1SM06570J} {\bibfield
  {journal} {\bibinfo  {journal} {Soft Matter}\ }\textbf {\bibinfo {volume}
  {8}},\ \bibinfo {pages} {667--672} (\bibinfo {year} {2012})}\BibitemShut
  {NoStop}%
\bibitem [{\citenamefont {Xia}\ \emph {et~al.}(2012)\citenamefont {Xia},
  \citenamefont {Nguyen}, \citenamefont {Yang}, \citenamefont {Lee},
  \citenamefont {Santos}, \citenamefont {Podsiadlo}, \citenamefont {Tang},
  \citenamefont {Glotzer},\ and\ \citenamefont {Kotov}}]{XiaGlotzer2012}%
  \BibitemOpen
  \bibfield  {author} {\bibinfo {author} {\bibfnamefont {Y.}~\bibnamefont
  {Xia}}, \bibinfo {author} {\bibfnamefont {T.~D.}\ \bibnamefont {Nguyen}},
  \bibinfo {author} {\bibfnamefont {M.}~\bibnamefont {Yang}}, \bibinfo {author}
  {\bibfnamefont {B.}~\bibnamefont {Lee}}, \bibinfo {author} {\bibfnamefont
  {A.}~\bibnamefont {Santos}}, \bibinfo {author} {\bibfnamefont
  {P.}~\bibnamefont {Podsiadlo}}, \bibinfo {author} {\bibfnamefont
  {Z.}~\bibnamefont {Tang}}, \bibinfo {author} {\bibfnamefont {S.~C.}\
  \bibnamefont {Glotzer}}, \ and\ \bibinfo {author} {\bibfnamefont {N.~A.}\
  \bibnamefont {Kotov}},\ }\bibfield  {title} {\enquote {\bibinfo {title}
  {Self-assembly of self-limiting monodisperse supraparticles from polydisperse
  nanoparticles},}\ }\href@noop {} {\bibfield  {journal} {\bibinfo  {journal}
  {Nat. Nanotechnol.}\ }\textbf {\bibinfo {volume} {7}},\ \bibinfo {pages}
  {479--479} (\bibinfo {year} {2012})}\BibitemShut {NoStop}%
\bibitem [{\citenamefont {Yethiraj}\ and\ \citenamefont {van
  Blaaderen}(2003)}]{Yethiraj2003}%
  \BibitemOpen
  \bibfield  {author} {\bibinfo {author} {\bibfnamefont {A.}~\bibnamefont
  {Yethiraj}}\ and\ \bibinfo {author} {\bibfnamefont {A.}~\bibnamefont {van
  Blaaderen}},\ }\bibfield  {title} {\enquote {\bibinfo {title} {A colloidal
  model system with an interaction tunable from hard sphere to soft and
  dipolar},}\ }\href {\doibase doi:10.1038/nature01328} {\bibfield  {journal}
  {\bibinfo  {journal} {Nature}\ }\textbf {\bibinfo {volume} {421}},\ \bibinfo
  {pages} {513--517} (\bibinfo {year} {2003})}\BibitemShut {NoStop}%
\bibitem [{\citenamefont {Stradner}\ \emph {et~al.}(2004)\citenamefont
  {Stradner}, \citenamefont {Sedgwick}, \citenamefont {Cardinaux},
  \citenamefont {Poon}, \citenamefont {Egelhaaf},\ and\ \citenamefont
  {Schurtenberger}}]{Stradner2004}%
  \BibitemOpen
  \bibfield  {author} {\bibinfo {author} {\bibfnamefont {A.}~\bibnamefont
  {Stradner}}, \bibinfo {author} {\bibfnamefont {H.}~\bibnamefont {Sedgwick}},
  \bibinfo {author} {\bibfnamefont {F.}~\bibnamefont {Cardinaux}}, \bibinfo
  {author} {\bibfnamefont {W.~C.~K.}\ \bibnamefont {Poon}}, \bibinfo {author}
  {\bibfnamefont {S.~U.}\ \bibnamefont {Egelhaaf}}, \ and\ \bibinfo {author}
  {\bibfnamefont {P.}~\bibnamefont {Schurtenberger}},\ }\bibfield  {title}
  {\enquote {\bibinfo {title} {Equilibrium cluster formation in concentrated
  protein solutions and colloids},}\ }\href {\doibase doi:10.1038/nature03109}
  {\bibfield  {journal} {\bibinfo  {journal} {Nature}\ }\textbf {\bibinfo
  {volume} {432}},\ \bibinfo {pages} {492--495} (\bibinfo {year}
  {2004})}\BibitemShut {NoStop}%
\bibitem [{\citenamefont {Porcar}\ \emph {et~al.}(2010)\citenamefont {Porcar},
  \citenamefont {Falus}, \citenamefont {Chen}, \citenamefont {Faraone},
  \citenamefont {Fratini}, \citenamefont {Hong}, \citenamefont {Baglioni},\
  and\ \citenamefont {Liu}}]{PorcarLiu2010}%
  \BibitemOpen
  \bibfield  {author} {\bibinfo {author} {\bibfnamefont {L.}~\bibnamefont
  {Porcar}}, \bibinfo {author} {\bibfnamefont {P.}~\bibnamefont {Falus}},
  \bibinfo {author} {\bibfnamefont {W.-R.}\ \bibnamefont {Chen}}, \bibinfo
  {author} {\bibfnamefont {A.}~\bibnamefont {Faraone}}, \bibinfo {author}
  {\bibfnamefont {E.}~\bibnamefont {Fratini}}, \bibinfo {author} {\bibfnamefont
  {K.}~\bibnamefont {Hong}}, \bibinfo {author} {\bibfnamefont {P.}~\bibnamefont
  {Baglioni}}, \ and\ \bibinfo {author} {\bibfnamefont {Y.}~\bibnamefont
  {Liu}},\ }\bibfield  {title} {\enquote {\bibinfo {title} {Formation of the
  dynamic clusters in concentrated lysozyme protein solutions},}\ }\href
  {\doibase 10.1021/jz900127c} {\bibfield  {journal} {\bibinfo  {journal} {J.
  Phys. Chem. Lett.}\ }\textbf {\bibinfo {volume} {1}},\ \bibinfo {pages}
  {126--129} (\bibinfo {year} {2010})},\ \Eprint
  {http://arxiv.org/abs/http://dx.doi.org/10.1021/jz900127c}
  {http://dx.doi.org/10.1021/jz900127c} \BibitemShut {NoStop}%
\bibitem [{\citenamefont {Liu}\ \emph {et~al.}(2011)\citenamefont {Liu},
  \citenamefont {Porcar}, \citenamefont {Chen}, \citenamefont {Chen},
  \citenamefont {Falus}, \citenamefont {Faraone}, \citenamefont {Fratini},
  \citenamefont {Hong},\ and\ \citenamefont {Baglioni}}]{LiuBaglioni2011}%
  \BibitemOpen
  \bibfield  {author} {\bibinfo {author} {\bibfnamefont {Y.}~\bibnamefont
  {Liu}}, \bibinfo {author} {\bibfnamefont {L.}~\bibnamefont {Porcar}},
  \bibinfo {author} {\bibfnamefont {J.}~\bibnamefont {Chen}}, \bibinfo {author}
  {\bibfnamefont {W.-R.}\ \bibnamefont {Chen}}, \bibinfo {author}
  {\bibfnamefont {P.}~\bibnamefont {Falus}}, \bibinfo {author} {\bibfnamefont
  {A.}~\bibnamefont {Faraone}}, \bibinfo {author} {\bibfnamefont
  {E.}~\bibnamefont {Fratini}}, \bibinfo {author} {\bibfnamefont
  {K.}~\bibnamefont {Hong}}, \ and\ \bibinfo {author} {\bibfnamefont
  {P.}~\bibnamefont {Baglioni}},\ }\bibfield  {title} {\enquote {\bibinfo
  {title} {Lysozyme protein solution with an intermediate range order
  structure},}\ }\href {\doibase 10.1021/jp109333c} {\bibfield  {journal}
  {\bibinfo  {journal} {J. Phys. Chem. B}\ }\textbf {\bibinfo {volume} {115}},\
  \bibinfo {pages} {7238--7247} (\bibinfo {year} {2011})},\ \Eprint
  {http://arxiv.org/abs/http://dx.doi.org/10.1021/jp109333c}
  {http://dx.doi.org/10.1021/jp109333c} \BibitemShut {NoStop}%
\bibitem [{\citenamefont {Johnston}\ \emph {et~al.}(2012)\citenamefont
  {Johnston}, \citenamefont {Maynard}, \citenamefont {Truskett}, \citenamefont
  {Borwankar}, \citenamefont {Miller}, \citenamefont {Wilson}, \citenamefont
  {Dinin}, \citenamefont {Khan},\ and\ \citenamefont
  {Kaczorowski}}]{Johnston2012}%
  \BibitemOpen
  \bibfield  {author} {\bibinfo {author} {\bibfnamefont {K.~P.}\ \bibnamefont
  {Johnston}}, \bibinfo {author} {\bibfnamefont {J.~A.}\ \bibnamefont
  {Maynard}}, \bibinfo {author} {\bibfnamefont {T.~M.}\ \bibnamefont
  {Truskett}}, \bibinfo {author} {\bibfnamefont {A.~U.}\ \bibnamefont
  {Borwankar}}, \bibinfo {author} {\bibfnamefont {M.~A.}\ \bibnamefont
  {Miller}}, \bibinfo {author} {\bibfnamefont {B.~K.}\ \bibnamefont {Wilson}},
  \bibinfo {author} {\bibfnamefont {A.~K.}\ \bibnamefont {Dinin}}, \bibinfo
  {author} {\bibfnamefont {T.~A.}\ \bibnamefont {Khan}}, \ and\ \bibinfo
  {author} {\bibfnamefont {K.~J.}\ \bibnamefont {Kaczorowski}},\ }\bibfield
  {title} {\enquote {\bibinfo {title} {Concentrated dispersions of equilibrium
  protein nanoclusters that reversibly dissociate into active monomers},}\
  }\href {\doibase 10.1021/nn204166z} {\bibfield  {journal} {\bibinfo
  {journal} {ACS Nano}\ }\textbf {\bibinfo {volume} {6}},\ \bibinfo {pages}
  {1357--1369} (\bibinfo {year} {2012})},\ \Eprint
  {http://arxiv.org/abs/http://dx.doi.org/10.1021/nn204166z}
  {http://dx.doi.org/10.1021/nn204166z} \BibitemShut {NoStop}%
\bibitem [{\citenamefont {Park}\ \emph {et~al.}(2014)\citenamefont {Park},
  \citenamefont {Nguyen}, \citenamefont {de~Queir{\'o}s~Silveira},
  \citenamefont {Bahng}, \citenamefont {Srivastava}, \citenamefont {Zhao},
  \citenamefont {Sun}, \citenamefont {Zhang}, \citenamefont {Glotzer},\ and\
  \citenamefont {Kotov}}]{ParkGlotzer2012}%
  \BibitemOpen
  \bibfield  {author} {\bibinfo {author} {\bibfnamefont {J.~I.}\ \bibnamefont
  {Park}}, \bibinfo {author} {\bibfnamefont {T.~D.}\ \bibnamefont {Nguyen}},
  \bibinfo {author} {\bibfnamefont {G.}~\bibnamefont
  {de~Queir{\'o}s~Silveira}}, \bibinfo {author} {\bibfnamefont {J.~H.}\
  \bibnamefont {Bahng}}, \bibinfo {author} {\bibfnamefont {S.}~\bibnamefont
  {Srivastava}}, \bibinfo {author} {\bibfnamefont {G.}~\bibnamefont {Zhao}},
  \bibinfo {author} {\bibfnamefont {K.}~\bibnamefont {Sun}}, \bibinfo {author}
  {\bibfnamefont {P.}~\bibnamefont {Zhang}}, \bibinfo {author} {\bibfnamefont
  {S.~C.}\ \bibnamefont {Glotzer}}, \ and\ \bibinfo {author} {\bibfnamefont
  {N.~A.}\ \bibnamefont {Kotov}},\ }\bibfield  {title} {\enquote {\bibinfo
  {title} {Terminal supraparticle assemblies from similarly charged protein
  molecules and nanoparticles},}\ }\href@noop {} {\bibfield  {journal}
  {\bibinfo  {journal} {Nat. Commun.}\ }\textbf {\bibinfo {volume} {5}}
  (\bibinfo {year} {2014})}\BibitemShut {NoStop}%
\bibitem [{\citenamefont {Yearley}\ \emph {et~al.}(2014)\citenamefont
  {Yearley}, \citenamefont {Godfrin}, \citenamefont {Perevozchikova},
  \citenamefont {Zhang}, \citenamefont {Falus}, \citenamefont {Porcar},
  \citenamefont {Nagao}, \citenamefont {Curtis}, \citenamefont {Gawande},
  \citenamefont {Taing}, \citenamefont {Zarraga}, \citenamefont {Wagner},\ and\
  \citenamefont {Liu}}]{Yearley2014}%
  \BibitemOpen
  \bibfield  {author} {\bibinfo {author} {\bibfnamefont {E.~J.}\ \bibnamefont
  {Yearley}}, \bibinfo {author} {\bibfnamefont {P.~D.}\ \bibnamefont
  {Godfrin}}, \bibinfo {author} {\bibfnamefont {T.}~\bibnamefont
  {Perevozchikova}}, \bibinfo {author} {\bibfnamefont {H.}~\bibnamefont
  {Zhang}}, \bibinfo {author} {\bibfnamefont {P.}~\bibnamefont {Falus}},
  \bibinfo {author} {\bibfnamefont {L.}~\bibnamefont {Porcar}}, \bibinfo
  {author} {\bibfnamefont {M.}~\bibnamefont {Nagao}}, \bibinfo {author}
  {\bibfnamefont {J.~E.}\ \bibnamefont {Curtis}}, \bibinfo {author}
  {\bibfnamefont {P.}~\bibnamefont {Gawande}}, \bibinfo {author} {\bibfnamefont
  {R.}~\bibnamefont {Taing}}, \bibinfo {author} {\bibfnamefont {I.~E.}\
  \bibnamefont {Zarraga}}, \bibinfo {author} {\bibfnamefont {N.~J.}\
  \bibnamefont {Wagner}}, \ and\ \bibinfo {author} {\bibfnamefont
  {Y.}~\bibnamefont {Liu}},\ }\bibfield  {title} {\enquote {\bibinfo {title}
  {Observation of small cluster formation in concentrated monoclonal antibody
  solutions and its implications to solution viscosity},}\ }\href {\doibase
  http://dx.doi.org/10.1016/j.bpj.2014.02.036} {\bibfield  {journal} {\bibinfo
  {journal} {Biophys. J.}\ }\textbf {\bibinfo {volume} {106}},\ \bibinfo
  {pages} {1763 -- 1770} (\bibinfo {year} {2014})}\BibitemShut {NoStop}%
\bibitem [{\citenamefont {Godfrin}\ \emph {et~al.}(2016)\citenamefont
  {Godfrin}, \citenamefont {Zarraga}, \citenamefont {Zarzar}, \citenamefont
  {Porcar}, \citenamefont {Falus}, \citenamefont {Wagner},\ and\ \citenamefont
  {Liu}}]{Godfrin2016}%
  \BibitemOpen
  \bibfield  {author} {\bibinfo {author} {\bibfnamefont {P.~D.}\ \bibnamefont
  {Godfrin}}, \bibinfo {author} {\bibfnamefont {I.~E.}\ \bibnamefont
  {Zarraga}}, \bibinfo {author} {\bibfnamefont {J.}~\bibnamefont {Zarzar}},
  \bibinfo {author} {\bibfnamefont {L.}~\bibnamefont {Porcar}}, \bibinfo
  {author} {\bibfnamefont {P.}~\bibnamefont {Falus}}, \bibinfo {author}
  {\bibfnamefont {N.~J.}\ \bibnamefont {Wagner}}, \ and\ \bibinfo {author}
  {\bibfnamefont {Y.}~\bibnamefont {Liu}},\ }\bibfield  {title} {\enquote
  {\bibinfo {title} {Effect of hierarchical cluster formation on the viscosity
  of concentrated monoclonal antibody formulations studied by neutron
  scattering},}\ }\href {\doibase 10.1021/acs.jpcb.5b07260} {\bibfield
  {journal} {\bibinfo  {journal} {J. Phys. Chem. B}\ }\textbf {\bibinfo
  {volume} {120}},\ \bibinfo {pages} {278--291} (\bibinfo {year} {2016})},\
  \bibinfo {note} {pMID: 26707135},\ \Eprint
  {http://arxiv.org/abs/http://dx.doi.org/10.1021/acs.jpcb.5b07260}
  {http://dx.doi.org/10.1021/acs.jpcb.5b07260} \BibitemShut {NoStop}%
\bibitem [{\citenamefont {Groenewold}\ and\ \citenamefont
  {Kegel}(2004)}]{Groenewold2004}%
  \BibitemOpen
  \bibfield  {author} {\bibinfo {author} {\bibfnamefont {J.}~\bibnamefont
  {Groenewold}}\ and\ \bibinfo {author} {\bibfnamefont {W.~K.}\ \bibnamefont
  {Kegel}},\ }\bibfield  {title} {\enquote {\bibinfo {title} {Colloidal cluster
  phases, gelation and nuclear matter},}\ }\href
  {http://stacks.iop.org/0953-8984/16/i=42/a=006} {\bibfield  {journal}
  {\bibinfo  {journal} {J. Phys.: Cond. Matt.}\ }\textbf {\bibinfo {volume}
  {16}},\ \bibinfo {pages} {S4877} (\bibinfo {year} {2004})}\BibitemShut
  {NoStop}%
\bibitem [{\citenamefont {Debenedetti}(1997)}]{Debenedetti1996}%
  \BibitemOpen
  \bibfield  {author} {\bibinfo {author} {\bibfnamefont {P.~G.}\ \bibnamefont
  {Debenedetti}},\ }\href@noop {} {\emph {\bibinfo {title} {Metastable Liquids:
  Concepts and Principles}}}\ (\bibinfo  {publisher} {Princeton University
  Press},\ \bibinfo {address} {Princeton, NJ, USA},\ \bibinfo {year}
  {1997})\BibitemShut {NoStop}%
\bibitem [{\citenamefont {Auer}\ and\ \citenamefont
  {Frenkel}(2000)}]{Auer2000}%
  \BibitemOpen
  \bibfield  {author} {\bibinfo {author} {\bibfnamefont {S.}~\bibnamefont
  {Auer}}\ and\ \bibinfo {author} {\bibfnamefont {D.}~\bibnamefont {Frenkel}},\
  }\bibfield  {title} {\enquote {\bibinfo {title} {Prediction of absolute
  crystal-nucleation rate in hard-sphere colloids},}\ }\href {\doibase
  doi:10.1038/35059035} {\bibfield  {journal} {\bibinfo  {journal} {Nature}\
  }\textbf {\bibinfo {volume} {409}},\ \bibinfo {pages} {1020--1023} (\bibinfo
  {year} {2000})}\BibitemShut {NoStop}%
\bibitem [{\citenamefont {Sear}(2007)}]{Sear2007}%
  \BibitemOpen
  \bibfield  {author} {\bibinfo {author} {\bibfnamefont {R.~P.}\ \bibnamefont
  {Sear}},\ }\bibfield  {title} {\enquote {\bibinfo {title} {Nucleation: theory
  and applications to protein solutions and colloidal suspensions},}\ }\href
  {http://stacks.iop.org/0953-8984/19/i=3/a=033101} {\bibfield  {journal}
  {\bibinfo  {journal} {J. Phys.: Cond. Matt.}\ }\textbf {\bibinfo {volume}
  {19}},\ \bibinfo {pages} {033101} (\bibinfo {year} {2007})}\BibitemShut
  {NoStop}%
\bibitem [{\citenamefont {Mani}\ and\ \citenamefont
  {L\"owen}(2015)}]{Mani2015}%
  \BibitemOpen
  \bibfield  {author} {\bibinfo {author} {\bibfnamefont {E.}~\bibnamefont
  {Mani}}\ and\ \bibinfo {author} {\bibfnamefont {H.}~\bibnamefont {L\"owen}},\
  }\bibfield  {title} {\enquote {\bibinfo {title} {Effect of self-propulsion on
  equilibrium clustering},}\ }\href {\doibase 10.1103/PhysRevE.92.032301}
  {\bibfield  {journal} {\bibinfo  {journal} {Phys. Rev. E}\ }\textbf {\bibinfo
  {volume} {92}},\ \bibinfo {pages} {032301} (\bibinfo {year}
  {2015})}\BibitemShut {NoStop}%
\bibitem [{\citenamefont {Arkus}, \citenamefont {Manoharan},\ and\
  \citenamefont {Brenner}(2009)}]{Arkus2009}%
  \BibitemOpen
  \bibfield  {author} {\bibinfo {author} {\bibfnamefont {N.}~\bibnamefont
  {Arkus}}, \bibinfo {author} {\bibfnamefont {V.~N.}\ \bibnamefont
  {Manoharan}}, \ and\ \bibinfo {author} {\bibfnamefont {M.~P.}\ \bibnamefont
  {Brenner}},\ }\bibfield  {title} {\enquote {\bibinfo {title} {Minimal energy
  clusters of hard spheres with short range attractions},}\ }\href {\doibase
  10.1103/PhysRevLett.103.118303} {\bibfield  {journal} {\bibinfo  {journal}
  {Phys. Rev. Lett.}\ }\textbf {\bibinfo {volume} {103}},\ \bibinfo {pages}
  {118303} (\bibinfo {year} {2009})}\BibitemShut {NoStop}%
\bibitem [{\citenamefont {Meng}\ \emph {et~al.}(2010)\citenamefont {Meng},
  \citenamefont {Arkus}, \citenamefont {Brenner},\ and\ \citenamefont
  {Manoharan}}]{Meng2010}%
  \BibitemOpen
  \bibfield  {author} {\bibinfo {author} {\bibfnamefont {G.}~\bibnamefont
  {Meng}}, \bibinfo {author} {\bibfnamefont {N.}~\bibnamefont {Arkus}},
  \bibinfo {author} {\bibfnamefont {M.~P.}\ \bibnamefont {Brenner}}, \ and\
  \bibinfo {author} {\bibfnamefont {V.~N.}\ \bibnamefont {Manoharan}},\
  }\bibfield  {title} {\enquote {\bibinfo {title} {The free-energy landscape of
  clusters of attractive hard spheres},}\ }\href {\doibase
  10.1126/science.1181263} {\bibfield  {journal} {\bibinfo  {journal}
  {Science}\ }\textbf {\bibinfo {volume} {327}},\ \bibinfo {pages} {560--563}
  (\bibinfo {year} {2010})},\ \Eprint
  {http://arxiv.org/abs/http://science.sciencemag.org/content/327/5965/560.full.pdf}
  {http://science.sciencemag.org/content/327/5965/560.full.pdf} \BibitemShut
  {NoStop}%
\bibitem [{\citenamefont {Jadrich}\ \emph
  {et~al.}(2015{\natexlab{b}})\citenamefont {Jadrich}, \citenamefont
  {Bollinger}, \citenamefont {Lindquist},\ and\ \citenamefont
  {Truskett}}]{JadrichSM2015}%
  \BibitemOpen
  \bibfield  {author} {\bibinfo {author} {\bibfnamefont {R.~B.}\ \bibnamefont
  {Jadrich}}, \bibinfo {author} {\bibfnamefont {J.~A.}\ \bibnamefont
  {Bollinger}}, \bibinfo {author} {\bibfnamefont {B.~A.}\ \bibnamefont
  {Lindquist}}, \ and\ \bibinfo {author} {\bibfnamefont {T.~M.}\ \bibnamefont
  {Truskett}},\ }\bibfield  {title} {\enquote {\bibinfo {title} {Equilibrium
  cluster fluids: pair interactions via inverse design},}\ }\href {\doibase
  10.1039/C5SM01832C} {\bibfield  {journal} {\bibinfo  {journal} {Soft Matter}\
  }\textbf {\bibinfo {volume} {11}},\ \bibinfo {pages} {9342--9354} (\bibinfo
  {year} {2015}{\natexlab{b}})}\BibitemShut {NoStop}%
\bibitem [{\citenamefont {Pandav}\ \emph {et~al.}(2015)\citenamefont {Pandav},
  \citenamefont {Pryamitsyn}, \citenamefont {Errington},\ and\ \citenamefont
  {Ganesan}}]{PandavJPCB2015}%
  \BibitemOpen
  \bibfield  {author} {\bibinfo {author} {\bibfnamefont {G.}~\bibnamefont
  {Pandav}}, \bibinfo {author} {\bibfnamefont {V.}~\bibnamefont {Pryamitsyn}},
  \bibinfo {author} {\bibfnamefont {J.}~\bibnamefont {Errington}}, \ and\
  \bibinfo {author} {\bibfnamefont {V.}~\bibnamefont {Ganesan}},\ }\bibfield
  {title} {\enquote {\bibinfo {title} {Multibody interactions, phase behavior,
  and clustering in nanoparticle–polyelectrolyte mixtures},}\ }\href
  {\doibase 10.1021/acs.jpcb.5b07905} {\bibfield  {journal} {\bibinfo
  {journal} {J. Phys. Chem. B}\ }\textbf {\bibinfo {volume} {119}},\ \bibinfo
  {pages} {14536--14550} (\bibinfo {year} {2015})},\ \bibinfo {note} {pMID:
  26473468},\ \Eprint
  {http://arxiv.org/abs/http://dx.doi.org/10.1021/acs.jpcb.5b07905}
  {http://dx.doi.org/10.1021/acs.jpcb.5b07905} \BibitemShut {NoStop}%
\bibitem [{\citenamefont {Pandav}, \citenamefont {Pryamitsyn},\ and\
  \citenamefont {Ganesan}(2015)}]{PandavLang2015}%
  \BibitemOpen
  \bibfield  {author} {\bibinfo {author} {\bibfnamefont {G.}~\bibnamefont
  {Pandav}}, \bibinfo {author} {\bibfnamefont {V.}~\bibnamefont {Pryamitsyn}},
  \ and\ \bibinfo {author} {\bibfnamefont {V.}~\bibnamefont {Ganesan}},\
  }\bibfield  {title} {\enquote {\bibinfo {title} {Interactions and aggregation
  of charged nanoparticles in uncharged polymer solutions},}\ }\href {\doibase
  10.1021/acs.langmuir.5b02885} {\bibfield  {journal} {\bibinfo  {journal}
  {Langmuir}\ }\textbf {\bibinfo {volume} {31}},\ \bibinfo {pages}
  {12328--12338} (\bibinfo {year} {2015})},\ \bibinfo {note} {pMID: 26535914},\
  \Eprint {http://arxiv.org/abs/http://dx.doi.org/10.1021/acs.langmuir.5b02885}
  {http://dx.doi.org/10.1021/acs.langmuir.5b02885} \BibitemShut {NoStop}%
\bibitem [{\citenamefont {Manning}(1979)}]{Manning1979}%
  \BibitemOpen
  \bibfield  {author} {\bibinfo {author} {\bibfnamefont {G.~S.}\ \bibnamefont
  {Manning}},\ }\bibfield  {title} {\enquote {\bibinfo {title} {Counterion
  binding in polyelectrolyte theory},}\ }\href {\doibase 10.1021/ar50144a004}
  {\bibfield  {journal} {\bibinfo  {journal} {Acc. Chem. Res.}\ }\textbf
  {\bibinfo {volume} {12}},\ \bibinfo {pages} {443--449} (\bibinfo {year}
  {1979})},\ \Eprint
  {http://arxiv.org/abs/http://dx.doi.org/10.1021/ar50144a004}
  {http://dx.doi.org/10.1021/ar50144a004} \BibitemShut {NoStop}%
\bibitem [{\citenamefont {Alexander}\ \emph {et~al.}(1984)\citenamefont
  {Alexander}, \citenamefont {Chaikin}, \citenamefont {Grant}, \citenamefont
  {Morales}, \citenamefont {Pincus},\ and\ \citenamefont
  {Hone}}]{Alexander1984}%
  \BibitemOpen
  \bibfield  {author} {\bibinfo {author} {\bibfnamefont {S.}~\bibnamefont
  {Alexander}}, \bibinfo {author} {\bibfnamefont {P.~M.}\ \bibnamefont
  {Chaikin}}, \bibinfo {author} {\bibfnamefont {P.}~\bibnamefont {Grant}},
  \bibinfo {author} {\bibfnamefont {G.~J.}\ \bibnamefont {Morales}}, \bibinfo
  {author} {\bibfnamefont {P.}~\bibnamefont {Pincus}}, \ and\ \bibinfo {author}
  {\bibfnamefont {D.}~\bibnamefont {Hone}},\ }\bibfield  {title} {\enquote
  {\bibinfo {title} {Charge renormalization, osmotic pressure, and bulk modulus
  of colloidal crystals: Theory},}\ }\href {\doibase
  http://dx.doi.org/10.1063/1.446600} {\bibfield  {journal} {\bibinfo
  {journal} {J. Chem. Phys.}\ }\textbf {\bibinfo {volume} {80}},\ \bibinfo
  {pages} {5776--5781} (\bibinfo {year} {1984})}\BibitemShut {NoStop}%
\bibitem [{\citenamefont {Ramanathan}(1988)}]{Ramanathan1988}%
  \BibitemOpen
  \bibfield  {author} {\bibinfo {author} {\bibfnamefont {G.~V.}\ \bibnamefont
  {Ramanathan}},\ }\bibfield  {title} {\enquote {\bibinfo {title} {Counterion
  condensation in micellar and colloidal solutions},}\ }\href {\doibase
  http://dx.doi.org/10.1063/1.453837} {\bibfield  {journal} {\bibinfo
  {journal} {J. Chem. Phys.}\ }\textbf {\bibinfo {volume} {88}},\ \bibinfo
  {pages} {3887--3892} (\bibinfo {year} {1988})}\BibitemShut {NoStop}%
\bibitem [{\citenamefont {Gillespie}\ \emph {et~al.}(2014)\citenamefont
  {Gillespie}, \citenamefont {Hallett}, \citenamefont {Elujoba}, \citenamefont
  {Che~Hamzah}, \citenamefont {Richardson},\ and\ \citenamefont
  {Bartlett}}]{Gillespie2014}%
  \BibitemOpen
  \bibfield  {author} {\bibinfo {author} {\bibfnamefont {D.~A.~J.}\
  \bibnamefont {Gillespie}}, \bibinfo {author} {\bibfnamefont {J.~E.}\
  \bibnamefont {Hallett}}, \bibinfo {author} {\bibfnamefont {O.}~\bibnamefont
  {Elujoba}}, \bibinfo {author} {\bibfnamefont {A.~F.}\ \bibnamefont
  {Che~Hamzah}}, \bibinfo {author} {\bibfnamefont {R.~M.}\ \bibnamefont
  {Richardson}}, \ and\ \bibinfo {author} {\bibfnamefont {P.}~\bibnamefont
  {Bartlett}},\ }\bibfield  {title} {\enquote {\bibinfo {title} {Counterion
  condensation on spheres in the salt-free limit},}\ }\href {\doibase
  10.1039/C3SM52563E} {\bibfield  {journal} {\bibinfo  {journal} {Soft Matter}\
  }\textbf {\bibinfo {volume} {10}},\ \bibinfo {pages} {566--577} (\bibinfo
  {year} {2014})}\BibitemShut {NoStop}%
\bibitem [{\citenamefont {Plimpton}(1995)}]{Plimpton1995}%
  \BibitemOpen
  \bibfield  {author} {\bibinfo {author} {\bibfnamefont {S.}~\bibnamefont
  {Plimpton}},\ }\bibfield  {title} {\enquote {\bibinfo {title} {Fast parallel
  algorithms for short-range molecular dynamics},}\ }\href {\doibase
  http://dx.doi.org/10.1006/jcph.1995.1039} {\bibfield  {journal} {\bibinfo
  {journal} {J. Comput. Phys.}\ }\textbf {\bibinfo {volume} {117}},\ \bibinfo
  {pages} {1--19} (\bibinfo {year} {1995})}\BibitemShut {NoStop}%
\bibitem [{\citenamefont {Sear}(1999{\natexlab{a}})}]{SearJCP1999}%
  \BibitemOpen
  \bibfield  {author} {\bibinfo {author} {\bibfnamefont {R.~P.}\ \bibnamefont
  {Sear}},\ }\bibfield  {title} {\enquote {\bibinfo {title} {Classical
  nucleation theory for the nucleation of the solid phase of spherical
  particles with a short-ranged attraction},}\ }\href {\doibase
  http://dx.doi.org/10.1063/1.479512} {\bibfield  {journal} {\bibinfo
  {journal} {J. Chem. Phys.}\ }\textbf {\bibinfo {volume} {111}},\ \bibinfo
  {pages} {2001--2007} (\bibinfo {year} {1999}{\natexlab{a}})}\BibitemShut
  {NoStop}%
\bibitem [{\citenamefont {Sear}(1999{\natexlab{b}})}]{SearPRE1999}%
  \BibitemOpen
  \bibfield  {author} {\bibinfo {author} {\bibfnamefont {R.~P.}\ \bibnamefont
  {Sear}},\ }\bibfield  {title} {\enquote {\bibinfo {title} {Low-temperature
  interface between the gas and solid phases of hard spheres with a
  short-ranged attraction},}\ }\href {\doibase 10.1103/PhysRevE.59.6838}
  {\bibfield  {journal} {\bibinfo  {journal} {Phys. Rev. E}\ }\textbf {\bibinfo
  {volume} {59}},\ \bibinfo {pages} {6838--6841} (\bibinfo {year}
  {1999}{\natexlab{b}})}\BibitemShut {NoStop}%
\bibitem [{Note1()}]{Note1}%
  \BibitemOpen
  \bibinfo {note} {In principle, the free-energy penalty also includes an
  \protect \emph {entropic} contribution due to the increased mobility
  particles might have at the droplet surface compared to the droplet interior;
  however, this contribution is often negligible~\cite {SearPRE1999}.
  Groenewold and co-workers do not address this issue~\cite
  {GroenewoldKegel2001,Groenewold2004,Zhang2012}, but for our systems, where
  clusters possess fluid-like structures with frequent rearrangement between
  interior to exterior (nevermind frequent intercluster exchange), we also
  expect this entropic differential to be small.}\BibitemShut {Stop}%
\bibitem [{Note2()}]{Note2}%
  \BibitemOpen
  \bibinfo {note} {Zhang and co-workers~\cite {Zhang2012} report the wrong
  exponent with respect to \(N\) for this term.}\BibitemShut {Stop}%
\bibitem [{\citenamefont {Tolman}(1949)}]{Tolman1949}%
  \BibitemOpen
  \bibfield  {author} {\bibinfo {author} {\bibfnamefont {R.~C.}\ \bibnamefont
  {Tolman}},\ }\bibfield  {title} {\enquote {\bibinfo {title} {The effect of
  droplet size on surface tension},}\ }\href {\doibase
  http://dx.doi.org/10.1063/1.1747247} {\bibfield  {journal} {\bibinfo
  {journal} {J. Chem. Phys.}\ }\textbf {\bibinfo {volume} {17}},\ \bibinfo
  {pages} {333--337} (\bibinfo {year} {1949})}\BibitemShut {NoStop}%
\bibitem [{\citenamefont {Nijmeijer}\ \emph {et~al.}(1992)\citenamefont
  {Nijmeijer}, \citenamefont {Bruin}, \citenamefont {van Woerkom},
  \citenamefont {Bakker},\ and\ \citenamefont {van Leeuwen}}]{Nijmeijer1992}%
  \BibitemOpen
  \bibfield  {author} {\bibinfo {author} {\bibfnamefont {M.~J.~P.}\
  \bibnamefont {Nijmeijer}}, \bibinfo {author} {\bibfnamefont {C.}~\bibnamefont
  {Bruin}}, \bibinfo {author} {\bibfnamefont {A.~B.}\ \bibnamefont {van
  Woerkom}}, \bibinfo {author} {\bibfnamefont {A.~F.}\ \bibnamefont {Bakker}},
  \ and\ \bibinfo {author} {\bibfnamefont {J.~M.~J.}\ \bibnamefont {van
  Leeuwen}},\ }\bibfield  {title} {\enquote {\bibinfo {title} {Molecular
  dynamics of the surface tension of a drop},}\ }\href {\doibase
  http://dx.doi.org/10.1063/1.462495} {\bibfield  {journal} {\bibinfo
  {journal} {J. Chem. Phys.}\ }\textbf {\bibinfo {volume} {96}},\ \bibinfo
  {pages} {565--576} (\bibinfo {year} {1992})}\BibitemShut {NoStop}%
\bibitem [{\citenamefont {McGraw}\ and\ \citenamefont
  {Laaksonen}(1996)}]{McGraw1996}%
  \BibitemOpen
  \bibfield  {author} {\bibinfo {author} {\bibfnamefont {R.}~\bibnamefont
  {McGraw}}\ and\ \bibinfo {author} {\bibfnamefont {A.}~\bibnamefont
  {Laaksonen}},\ }\bibfield  {title} {\enquote {\bibinfo {title} {Scaling
  properties of the critical nucleus in classical and molecular-based theories
  of vapor-liquid nucleation},}\ }\href {\doibase 10.1103/PhysRevLett.76.2754}
  {\bibfield  {journal} {\bibinfo  {journal} {Phys. Rev. Lett.}\ }\textbf
  {\bibinfo {volume} {76}},\ \bibinfo {pages} {2754--2757} (\bibinfo {year}
  {1996})}\BibitemShut {NoStop}%
\bibitem [{\citenamefont {ten Wolde}\ and\ \citenamefont
  {Frenkel}(1998)}]{tenWolde1998}%
  \BibitemOpen
  \bibfield  {author} {\bibinfo {author} {\bibfnamefont {P.~R.}\ \bibnamefont
  {ten Wolde}}\ and\ \bibinfo {author} {\bibfnamefont {D.}~\bibnamefont
  {Frenkel}},\ }\bibfield  {title} {\enquote {\bibinfo {title} {Computer
  simulation study of gas-liquid nucleation in a {L}ennard-{J}ones system},}\
  }\href {\doibase http://dx.doi.org/10.1063/1.477658} {\bibfield  {journal}
  {\bibinfo  {journal} {J. Chem. Phys.}\ }\textbf {\bibinfo {volume} {109}},\
  \bibinfo {pages} {9901--9918} (\bibinfo {year} {1998})}\BibitemShut {NoStop}%
\bibitem [{\citenamefont {Koga}, \citenamefont {Zeng},\ and\ \citenamefont
  {Shchekin}(1998)}]{Koga1998}%
  \BibitemOpen
  \bibfield  {author} {\bibinfo {author} {\bibfnamefont {K.}~\bibnamefont
  {Koga}}, \bibinfo {author} {\bibfnamefont {X.~C.}\ \bibnamefont {Zeng}}, \
  and\ \bibinfo {author} {\bibfnamefont {A.~K.}\ \bibnamefont {Shchekin}},\
  }\bibfield  {title} {\enquote {\bibinfo {title} {Validity of {T}olman's
  equation: {H}ow large should a droplet be?}}\ }\href {\doibase
  http://dx.doi.org/10.1063/1.477006} {\bibfield  {journal} {\bibinfo
  {journal} {J. Chem. Phys.}\ }\textbf {\bibinfo {volume} {109}},\ \bibinfo
  {pages} {4063--4070} (\bibinfo {year} {1998})}\BibitemShut {NoStop}%
\bibitem [{\citenamefont {van Giessen}\ and\ \citenamefont
  {Blokhuis}(2009)}]{vanGiessen2009}%
  \BibitemOpen
  \bibfield  {author} {\bibinfo {author} {\bibfnamefont {A.~E.}\ \bibnamefont
  {van Giessen}}\ and\ \bibinfo {author} {\bibfnamefont {E.~M.}\ \bibnamefont
  {Blokhuis}},\ }\bibfield  {title} {\enquote {\bibinfo {title} {Direct
  determination of the {T}olman length from the bulk pressures of liquid drops
  via molecular dynamics simulations},}\ }\href {\doibase
  http://dx.doi.org/10.1063/1.3253685} {\bibfield  {journal} {\bibinfo
  {journal} {J. Chem. Phys.}\ }\textbf {\bibinfo {volume} {131}},\ \bibinfo
  {eid} {164705} (\bibinfo {year} {2009})}\BibitemShut {NoStop}%
\bibitem [{\citenamefont {Tr{\"o}ster}\ \emph {et~al.}(2012)\citenamefont
  {Tr{\"o}ster}, \citenamefont {Oettel}, \citenamefont {Block}, \citenamefont
  {Virnau},\ and\ \citenamefont {Binder}}]{Troster2012}%
  \BibitemOpen
  \bibfield  {author} {\bibinfo {author} {\bibfnamefont {A.}~\bibnamefont
  {Tr{\"o}ster}}, \bibinfo {author} {\bibfnamefont {M.}~\bibnamefont {Oettel}},
  \bibinfo {author} {\bibfnamefont {B.}~\bibnamefont {Block}}, \bibinfo
  {author} {\bibfnamefont {P.}~\bibnamefont {Virnau}}, \ and\ \bibinfo {author}
  {\bibfnamefont {K.}~\bibnamefont {Binder}},\ }\bibfield  {title} {\enquote
  {\bibinfo {title} {Numerical approaches to determine the interface tension of
  curved interfaces from free energy calculations},}\ }\href {\doibase
  http://dx.doi.org/10.1063/1.3685221} {\bibfield  {journal} {\bibinfo
  {journal} {J. Chem. Phys.}\ }\textbf {\bibinfo {volume} {136}},\ \bibinfo
  {eid} {064709} (\bibinfo {year} {2012})}\BibitemShut {NoStop}%
\bibitem [{\citenamefont {Wilhelmsen}, \citenamefont {Bedeaux},\ and\
  \citenamefont {Reguera}(2015)}]{Wilhelmsen2015}%
  \BibitemOpen
  \bibfield  {author} {\bibinfo {author} {\bibfnamefont {{\O}.}~\bibnamefont
  {Wilhelmsen}}, \bibinfo {author} {\bibfnamefont {D.}~\bibnamefont {Bedeaux}},
  \ and\ \bibinfo {author} {\bibfnamefont {D.}~\bibnamefont {Reguera}},\
  }\bibfield  {title} {\enquote {\bibinfo {title} {{T}olman length and rigidity
  constants of the {L}ennard-{J}ones fluid},}\ }\href {\doibase
  http://dx.doi.org/10.1063/1.4907588} {\bibfield  {journal} {\bibinfo
  {journal} {J. Chem. Phys.}\ }\textbf {\bibinfo {volume} {142}},\ \bibinfo
  {eid} {064706} (\bibinfo {year} {2015})}\BibitemShut {NoStop}%
\bibitem [{Note3()}]{Note3}%
  \BibitemOpen
  \bibinfo {note} {Note that the prefactor \(k\) modestly decreases as \(\phi
  \) increases: this occurs because, as discussed earlier, the cluster radius
  modestly increases with \(\phi \) for fixed \(N^{*}\); thus, clusters become
  less dense and exhibit correspondingly fewer bonds.}\BibitemShut {Stop}%
\bibitem [{\citenamefont {Pfender}\ and\ \citenamefont
  {Ziegler}(2004)}]{Pfender2004}%
  \BibitemOpen
  \bibfield  {author} {\bibinfo {author} {\bibfnamefont {F.}~\bibnamefont
  {Pfender}}\ and\ \bibinfo {author} {\bibfnamefont {G.~M.}\ \bibnamefont
  {Ziegler}},\ }\bibfield  {title} {\enquote {\bibinfo {title} {Kissing
  numbers, sphere packings and some unexpected proofs},}\ }\href@noop {}
  {\bibfield  {journal} {\bibinfo  {journal} {Notices Amer. Math. Soc}\
  }\textbf {\bibinfo {volume} {51}},\ \bibinfo {pages} {873--883} (\bibinfo
  {year} {2004})}\BibitemShut {NoStop}%
\bibitem [{Note4()}]{Note4}%
  \BibitemOpen
  \bibinfo {note} {The relation between coordination number and cluster size
  that we observe, \(z_{\protect \text {c}}(N) = k\protect \qopname \relax
  o{ln}(N)\) (with \(k \approx 2\)), has a much stronger scaling than that of a
  similar relation reported by Godfrin et. al.~\cite {GodfrinWagnerLiu2014},
  which was given as \(z_{\protect \text {c}}(N) = 1.5[\protect \qopname \relax
  o{ln}(N)]^{1/2}\) (here written in our choice of notation). We would simply
  note that the latter reaches an \protect \emph {asymptotic} coordination
  number of approximately 4 at very large droplet sizes, which would point to
  extremely elongated non-compact clusters (even Bernal spiral motifs~\cite
  {Campbell2005} exhibit \(z_{\protect \text {c}} \approx 5\)). In contrast,
  our expression, which is based on data from compact spherical aggregates at
  the onset of clustering, grows with cluster size and tends to approach the
  bulk coordination number \(z_{\protect \text {bulk}}=12\) of a dense
  attractive fluid in the large \(N\) limit, as in Fig. 5(b).}\BibitemShut
  {Stop}%
\bibitem [{Note5()}]{Note5}%
  \BibitemOpen
  \bibinfo {note} {In our approximate treatment, we suspect that \protect \emph
  {at small} \(N\), we are simultaneously: (1) underestimating the relative
  fraction of surface particles, which means the number of ``surface''
  particles actually scales as \(N^{m}\) with \(m<2/3\) over the whole
  intermediate size range; and (2) overestimating the coordination number
  \(z_{\protect \text {c}}(N)\) of surface particles (underestimating
  \(z_{\protect \text {c,m}}(N)\)), which means that the number of missing
  surface bonds actually scales as \(z_{\protect \text {c,m}}(N) \propto
  N^{m}\) with \(m>-1/3\). Because these errors in the exponents tend to cancel
  each other, we expect that the net effective \(N^{1/3}\) scaling of the
  surface term in Eqn.~\ref {eqn:reqFsurf} holds even given greater precision
  in the configurational analysis.}\BibitemShut {Stop}%
\end{thebibliography}


\end{document}